\documentclass[10pt,conference]{IEEEtran}

\usepackage[ruled,vlined]{algorithm2e}
\usepackage{graphicx,dblfloatfix}
\usepackage{url}
\usepackage{cite}
\usepackage[center]{subfigure}
\usepackage{times}
\usepackage{amsmath, amssymb,latexsym}
\usepackage{enumerate}
\usepackage{epsfig}
\usepackage{epsfig,epstopdf}
\usepackage[super]{nth}
\usepackage{color}
\usepackage{verbatim}

\newif\ifremark
\long\def\remark#1{
\ifremark%
	\begingroup%
	\dimen0=\columnwidth
	\advance\dimen0 by -0.25in%
	\setbox0=\hbox{\parbox[b]{\dimen0}{\protect\em #1}}
	\dimen1=\ht0\advance\dimen1 by 2pt%
	\dimen2=\dp0\advance\dimen2 by 2pt%
	\vskip 0.25pt%
	\hbox to \columnwidth{%
		\vrule height\dimen1 width 3pt depth\dimen2%
		\hss\copy0\hss%
		\vrule height\dimen1 width 3pt depth\dimen2%
	}%
	\endgroup%
\fi}

\remarktrue

\providecommand{\eat}[1]{}

\makeatletter
\newcommand{\removelatexerror}{\let\@latex@error\@gobble}
\makeatother
\begin{document}
\title{Coordinated Dynamic Spectrum Management of LTE-U and Wi-Fi Networks}
\author{Shweta Sagari\authorrefmark{1}, Samuel Baysting\authorrefmark{1}, Dola Saha\authorrefmark{2}, Ivan Seskar\authorrefmark{1}, Wade Trappe\authorrefmark{1}, Dipankar Raychaudhuri\authorrefmark{1}\\
\authorrefmark{1}WINLAB, Rutgers University, {\em\{shsagari, sbaysting, seskar, trappe, ray\}@winlab.rutgers.edu}
\\ \authorrefmark{2}NEC Labs America, {\em dola@nec-labs.com}
}

\maketitle
\thispagestyle{empty}
\let\VERBATIM\verbatim
\def\verbatim{%
\def\verbatim@font{\small\ttfamily}%
\VERBATIM}

\begin{abstract}
This paper investigates the co-existence of Wi-Fi and LTE in emerging unlicensed frequency bands which are intended to accommodate multiple radio access technologies.  Wi-Fi and LTE are the two most prominent access technologies being deployed today, motivating further study of the inter-system interference arising in such shared spectrum scenarios as well as possible techniques for enabling improved co-existence.  An analytical model for evaluating the baseline performance of co-existing Wi-Fi and LTE is developed and used to obtain baseline performance measures.  The results show that both Wi-Fi and LTE networks cause significant interference to each other and that the degradation is dependent on a number of factors such as power levels and physical topology.  The model-based results are partially validated via experimental evaluations using USRP based SDR platforms on the ORBIT testbed.  Further, inter-network coordination with logically centralized radio resource management across Wi-Fi and LTE systems is proposed as a possible solution for improved co-existence.  Numerical results are presented showing significant gains in both Wi-Fi and LTE performance with the proposed inter-network coordination approach.
\end{abstract}

\section{Introduction}
\label{sec:intro}
Exponential growth in mobile data usage is driven by the fact that Internet applications of all kinds are rapidly migrating from wired PCs to mobile smartphones, tablets, mobile APs and other portable devices~\cite{Cisco2014_VNI}. Industry has already started gearing up for the 1000x increase in data capacity, which has given rise to the concept of the 5th Generation (5G) mobile network. The 5G vision, though, is not limited to matching the increase in mobile data demand, but it also includes an improved overall service-oriented user experience with immersive applications, such as high definition video streaming, real-time interactive games, applications in wearable mobile devices, ubiquitous health care, mobile cloud, etc.~\cite{samsung2014_5G, giri2014_5G, Etemad2014_5G}. For such applications, the system needs to provide improved Quality of Experience (QoE), which can be factored in different ways: better cell/edge coverage (availability of service), lower latency (round trip time), lower power consumption (longer battery life), reliable services, cost-effective network, and support for mobility.

\begin{figure}[~t]
\begin{center}
\includegraphics[width=0.9\linewidth]{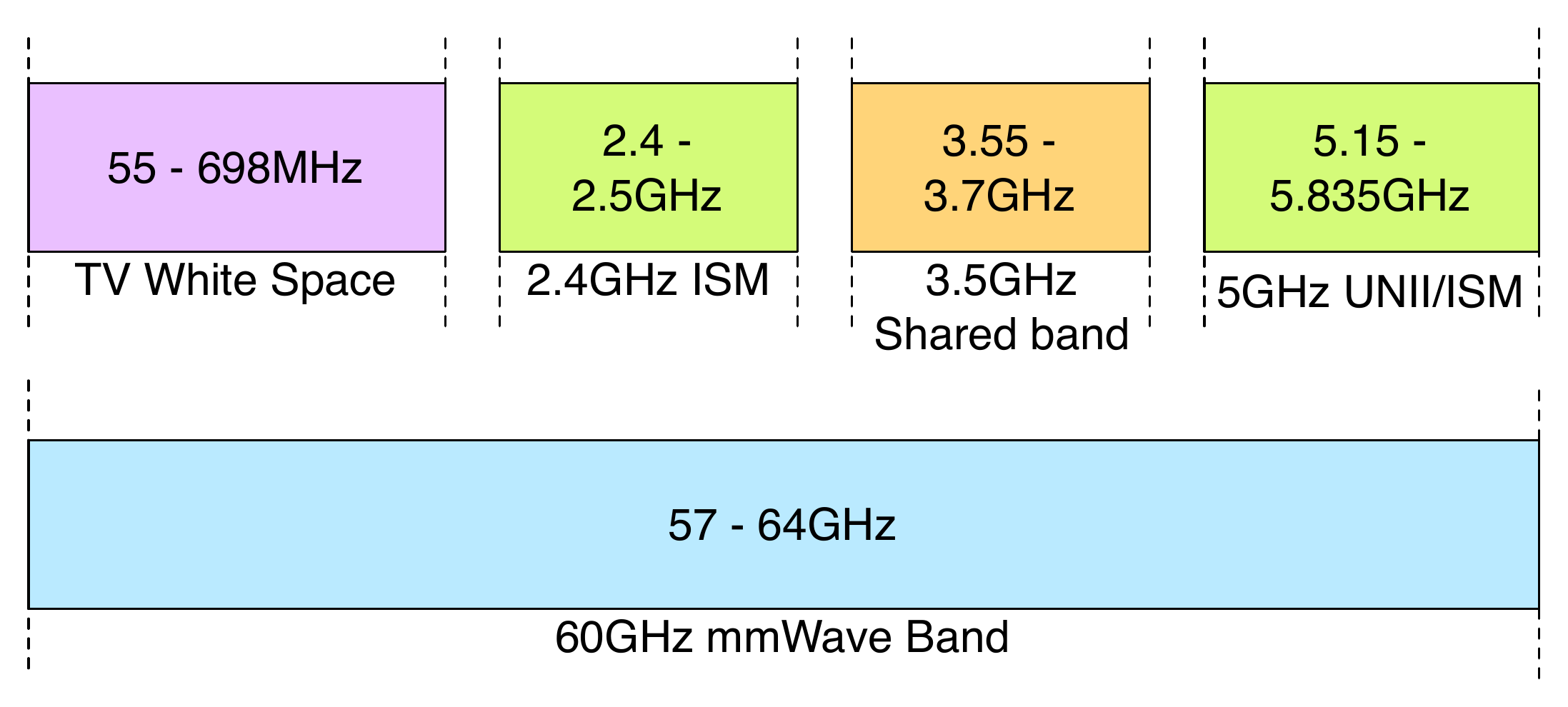}
\caption{Proposed spectrum bands for deployment of LTE/Wi-Fi small cells.}
\label{fig:spectBand}
\end{center}
\end{figure}

To meet such a high Quality-of-Service and system capacity demand, there have been three main solutions  proposed~\cite{Agilent2014_5G}: a) addition of more radio spectrum for mobile services (increase in MHz), b) deployment of small cells (increase in bits/Hz/km$^2$), and c) efficient spectrum utilization (increase in bits/second /Hz/km$^2$).
Several spectrum bands, as shown in figure \ref{fig:spectBand}, have been opened up for mobile and fixed wireless broadband services. These include 2.4 and 5 GHz unlicensed bands for the proposed unlicensed LTE operation as a secondary LTE carrier\cite{RP-140060}. These bands are currently utilized by unlicensed technologies such Wi-Fi/Bluetooth. The 3.5 GHz band, which is currently utilized for military and satellite operations has also been proposed for small cell (Wi-Fi/LTE based) services. Another possibility is the 60 GHz band (millimeter wave technology), which is well suited for short-distance communications including Gbps Wi-Fi, 5G cellular and peer-to-peer communications\cite{Sadri2013_mmWave}. In addition, opportunistic spectrum access is also possible in TV white spaces for small cell/backhaul operations as secondary users\cite{Shellhammer2009_techChallenges}.

\begin{figure}[!t]
\begin{center}
\subfigure[Interference caused by LTE.]
{\label{fig:lteInterferer}\includegraphics[width=0.45\linewidth]{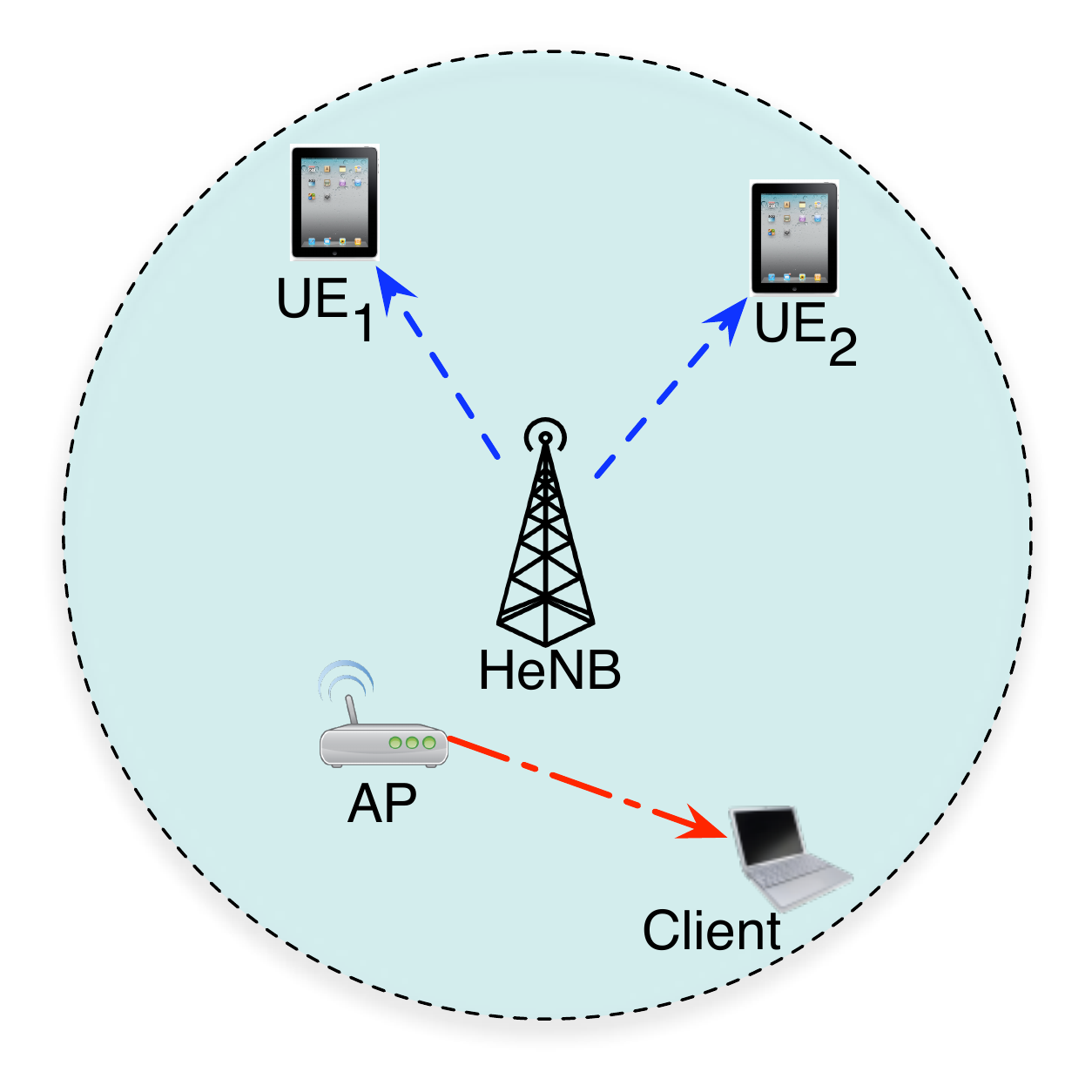}}
\subfigure[Interference caused by Wi-Fi.]
{\label{fig:wifiInterferer}\includegraphics[width=0.5\linewidth]{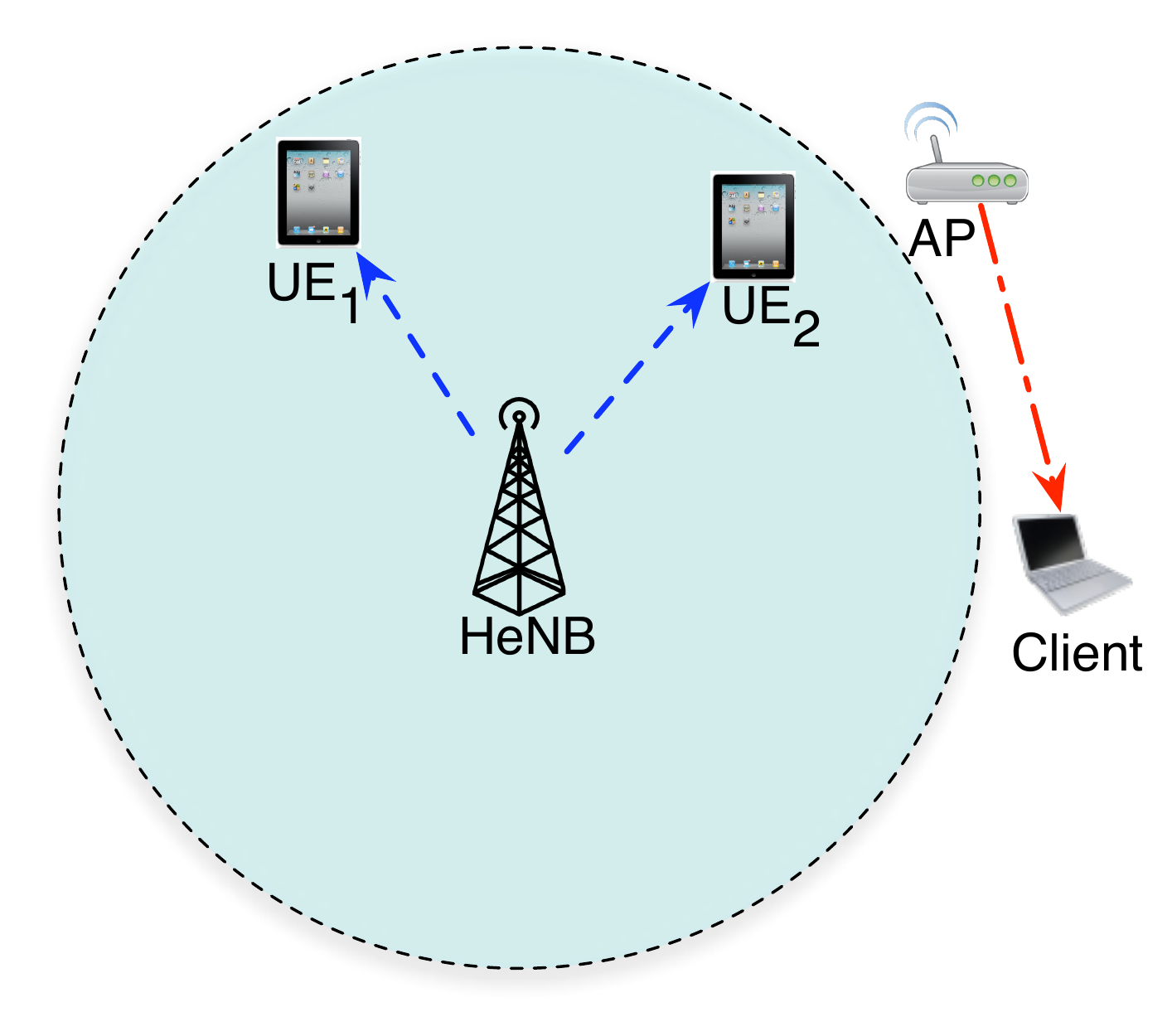}}
\caption{Scenarios showing challenges of coexistence of LTE and Wi-Fi in the same unlicensed spectrum.}
\label{fig:coexistChallenge}
\end{center}
\end{figure}

These emerging unlicensed band scenarios will lead to co-channel deployment of multiple radio access technologies (RATs) by multiple operators. These different RATs, designed for specific purposes at different frequencies, now must coexist in the same frequency, time and space. This causes increased interference to each other and degradation of the overall system performance is eminent due to the lack of inter-RAT compatibility. 
Figure \ref{fig:coexistChallenge} shows two such scenarios, where the two networks interfere with each other. When Wi-Fi Access Point is within the transmission zone of LTE, it senses the medium and postpones transmission due to detection of LTE Home eNodeB's (HeNB) transmission power in the spectrum band as shown in figure \ref{fig:lteInterferer}. Consequently, the Wi-Fi link from AP to Client suffers in presence of LTE transmission. The main reason for this disproportionate share of the medium is due to the fact that LTE does not sense other transmissions before transmitting. On the other hand, Wi-Fi is designed to coexist with other networks as it senses the channel before any transmission. However, if LTE works as supplemental downlink only mode, UEs do not transmit at all. So, a Wi-Fi AP, which cannot sense LTE HeNB's transmission, will transmit and cause interference at the nearby UEs, as shown in figure \ref{fig:wifiInterferer}. This problem also exists in multiple Wi-Fi links with some overlap in collision domain, but the network can recover packets quickly as a) packets are transmitted for a very short duration in Wi-Fi, compared to longer frames in LTE and b) all the nodes perform carrier sensing before transmission. 
Therefore, to fully utilize the benefits of new spectrum bands and deployments of HetNets, efficient spectrum utilization needs to be provided by the dynamic spectrum coordination framework and the supporting network architecture.

It is reasonable to forecast that Wi-Fi and LTE will be among the dominant technologies used by RATs for access purposes over the next few years. Thus, this paper focuses on the coordinated coexistence between these two technologies. LTE is designed to operate solely in a spectrum, which when operating in unlicensed spectrum, is termed LTE-U. It is suggested in 3GPP, that LTE-U will be used as a supplemental downlink, whereas the uplink will use licensed spectrum. This makes the deployment even more challenging as the UE's do not transmit in unlicensed spectrum yet experience interference from Wi-Fi transmissions. To alleviate these problems, we extend the interference characterization of co-channel deployment of Wi-Fi and LTE using simplistic but accurate analytical model\cite{Sagari2015_model}.  Then, we validate this model through experimental analysis of co-channel deployment in the 2.4 GHz band, using the ORBIT testbed and LTE on USRP platforms available at WINLAB.

To support the co-existence of a multi-RAT network, we propose a dynamic spectrum coordination framework, which is enabled by a Software Defined Network (SDN) architecture. SDN is technology-agnostic, can accommodate different radio standards and does not require change to existing standards or protocols. In contrast to existing technology-centric solutions, this is a desirable feature, especially in the rapid development of upcoming technologies and spectrum bands\cite{Raychaudhuri2014_NASCOR, Nychis2014_centralized}. Furthermore, the proposed framework takes advantage of the ubiquitous Internet connectivity available at wireless devices and provides the pseudo-global network with the ability to consider policy requirements in conjunction with improved visibility of each of the technologies, spectrum bands, clients and/or operators. Thus, it offers significant benefits for spectrum allocation over centralized spectrum servers\cite{Ileri2005_damand} or radio based control channels\cite{Jing2005_SpectrumCRMONET}.

While the inter-network cooperation enabled by SDN can be used for optimizing several spectrum usage parameters such as power control, channel selection, rate allocation, and duty cycle, in this paper, we focus on power control at both LTE and Wi-Fi, which maximizes aggregate throughput at all clients across both Wi-Fi and LTE networks along with consideration of throughput requirement at each client~\cite{Baid2015_understanding, Sagari2014_coexistence}. We also propose to apply validated interference characterization of Wi-Fi-LTE coexistence in the optimization framework, which captures the specific requirements of each of the technologies. In general, we adopt the geometric programming framework developed in \cite{Chiang2007_pwrCtr} for the LTE-only network and enhance it to accommodate Wi-Fi network.

The major contributions of this work are as follows:
\begin{itemize}
\item We introduce an analytical model to characterize the interference between Wi-Fi and LTE networks, when they coexist and share the medium in time, frequency and space. We have also validated the model by performing experimental analysis using USRP based LTE nodes and commercial off-the-shelf (COTS) IEEE 802.11g devices in the ORBIT testbed.
\item We propose a coordination framework to facilitate dynamic spectrum management among multi-operator and multi-technology networks over a large geographical area.
\item We propose a logically centralized cooperative optimization framework that involves dynamic coordination between Wi-Fi and LTE networks by exploiting power control and time division channel access diversity.
\item We evaluate the proposed optimization framework for improved coexistence between Wi-Fi and LTE networks.
\end{itemize}

The rest of the paper is organized as follows. In \S \ref{sec:related}, we discuss previous work on this topic and distinguish our work from existing literature. In \S \ref{sec:interChara}, we propose an analytical model to characterize the interference between Wi-Fi and LTE networks followed by partial experimental validation of the model. In \S \ref{sec:arch}, we propose an SDN-based inter-network coordination architecture, which can be used for transferring control messages between the different entities in the network. We use two approaches - power control and channel access time sharing methods to jointly optimize the spectrum sharing among Wi-Fi and LTE networks, which is described in \S \ref{sec:optimize}, followed by their evaluation in \S \ref{sec:results}. We conclude in \S \ref{sec:conclude}.

\section{Background on Wi-Fi/LTE Co-existence}
\label{sec:related}
Coordination between multi-RAT networks with LTE and Wi-Fi is challenging due to the difference in the medium access control layer of the two technologies.

Wi-Fi is based on the distributed coordination function (DCF) where each transmitter senses the channel energy for transmission opportunities and collision avoidance. In particular, clear channel assessment (CCA) in Wi-Fi involves two functions to detect any on-going transmissions\cite{IEEE_wlan,Jamieson2005_understand} -
\begin{enumerate}
\item \textit{Carrier sense}: Defines the ability of the Wi-Fi node to detect and decode other nodes' preambles, which most likely announces an incoming transmission. In such cases, Wi-Fi nodes are said to be in the CSMA range of each other other. For the basic DCF with no RTS/CTS, the Wi-Fi throughput can be accurately characterized using the Markov chain analysis given in Bianchi's model~\cite{bianchi2000_802.11DCF}, assuming a saturated traffic condition (at least 1 packet is waiting to be sent) at each node. Wi-Fi channel rates used in the \cite{bianchi2000_802.11DCF} can be modeled as a function of Signal-to-Interference-plus-Noise ratio. Our throughput analysis given in the following sections is based on Bianchi's model.
\item \textit{Energy detection}:
Defines the ability of Wi-Fi to detect non-Wi-Fi (in this case, LTE) energy in the operating channel and back off the data transmission. If the in-band signal energy crosses a certain threshold, the channel is detected as busy (no Wi-Fi transmission) until the channel energy is below the threshold. Thus, this function becomes the key parameter for characterizing Wi-Fi throughput in the co-channel deployment with LTE.
\end{enumerate}

LTE has both frequency division (FDD) and time division (TDD) multiplexing modes to operate. But to operate in unlicensed spectrum, supplemental downlink and TDD access is preferred. In either of the operations, data packets are scheduled in the successive time frames. LTE is based on orthogonal frequency-division multiple access (OFDMA), where a subset of subcarriers can be assigned to multiple users for a certain symbol time. This gives LTE additional diversity in the time and frequency domain that Wi-Fi lacks, since Wi-Fi bandwidth is assigned to a single user at any time. Further, LTE does not assume that spectrum is shared, and consequently does not employ any sharing features in the channel access mechanisms. Thus, the coexistence performance of both Wi-Fi and LTE is largely unpredictable and may lead to unfair spectrum sharing or the starvation of one of the technologies\cite{Chaves2013_LTEWiCoChallenges,Abinader2014_enablingLTEWiFi}.

In the literature, several studies have discussed spectrum management for multi-RAT heterogeneous networks in shared frequency bands, primarily focusing on IEEE 802.11 and 802.16 networks\cite{Jing2005_SpectrumCRMONET,Baid2011_spectrumMRI,Nychis2014_centralized}. Recently, Wi-Fi and LTE coexistence has been studied in the context of TV white space\cite{Almeida2013_EnaLTEWiFi}, in-device coexistence\cite{Baghel2011_coexiISMGNSS}, and LTE-unlicensed (LTE-U)\cite{NokiaVTC2013_perEval,Qualcomm2014_LTEHarmonious,Ratasuk2012_LicenseExempt}. Several studies \cite{Qualcomm2014_LTEHarmonious,Ratasuk2012_LicenseExempt,Liu2011_framework} propose CSMA/sensing based modifications in LTE with features like Listen-before-Talk, RTS/CTS protocol, and slotted channel access. In other studies, to enable Wi-Fi/LTE coexistence, solutions like blank LTE subframes/LTE muting (feature in LTE Release 10/11) \cite{Almeida2013_EnaLTEWiFi,Nihtila2013_SysPer}, carrier sensing adaptive transmission\cite{Qualcomm2014_LTEHarmonious}, interference aware power control in LTE\cite{Chaves2013_LTEULpower} have been proposed, which require LTE to transfer its resources to Wi-Fi. These schemes give Wi-Fi transmission opportunities but also lead to performance tradeoffs for LTE. Further, time domain solutions often require time synchronization between Wi-Fi and LTE and increase channel signaling. Some aspects of frequency and LTE bandwidth diversity have been explored in studies \cite{Qualcomm2014_LTEHarmonious} and \cite{Paiva2013_PHYLayer}, respectively. Frequency diversity is perhaps the least studied problem in Wi-Fi/LTE coexistence, while previous studies also have yet to consider dense Wi-Fi and LTE HetNet deployment scenarios in detail. Notably, in the literature, there are no previous studies experimentally evaluating the coexistence performance of Wi-Fi and LTE.

\eat{
In recent years, studies of Wi-Fi and LTE coexistence have been performed in the context of TV white space\cite{Almeida2013_EnaLTEWiFi}, in-device coexistence\cite{Baghel2011_coexiISMGNSS}, LTE-unlicensed (LTE-U)\cite{Qualcomm2014_LTEHarmonious,NokiaVTC2013_perEval}, etc. For the harmonious operation of LTE with Wi-Fi, Wi-Fi like feature Listen-before-Talk and Carrier-Sensing Adaptive Transmission has been proposed\cite{Qualcomm2014_LTEHarmonious}. Authors proposes slotted channel access based on channel sensing for LTE operation. To enable Wi-Fi/LTE coexistence, mechanisms like LTE blank subframe allocation (feature in LTE Release 10)\cite{Almeida2013_EnaLTEWiFi} and LTE uplink power control\cite{Chaves2013_LTEULpower} are been proposed to give opportunities for Wi-Fi transmission. Paiva et al. proposes to reduce LTE bandwidth to 5 MHz in the coexistence operation with Wi-Fi. Frequency diversity is the least studied problem in Wi-Fi/LTE coexistence. Also, previous studies have yet to consider dense Wi-Fi and LTE hetnet deployment scenarios in detail.
\cite{Ratasuk2012_LicenseExempt,Nihtila2013_SysPer}

\cite{Paiva2013_PHYLayer}: the a physical layer framework for the link-level coexistence simulation of LTE and Wi-Fi; both LTE and Wi-Fi have slightly larger BLER when the interference comes from the opposite technology. Additionally some of the LTE interference parameters effect on Wi-Fi performance was evaluated. It was shown that by reducing the LTE interference bandwidth from 20 MHz to 5 MHz the Wi-Fi results improve in SNR by 2.5 dB.

\cite{Liu2011_framework}: downlink operation of LTE for data only and all control signal are assumed to be communicated in licensed band. The proposed channel access mechanism of LTE in unlicensed band (Fig. 1 in the paper): Femtocell (FBS) can access channel only at the beginning of predefined time slots; At the beginning of time slot, if the channel is idle for Tsensing, then only it can use for maximum duration TcellTx; FBS cannot access channel at least for time duration of Tattempt between two consecutive transmission; the proposed mechanism is analyzed for a single Wi-Fi AP and single FBS coexistence through probabilistic model and simulation.

\cite{Chaves2013_LTEULpower}:LTE UL power control, controlled decrease of LTE UL transmit power according to inter-
ference measurements; proposed LTE UL power control can provide similar or better performance for both LTE and Wi-
Fi networks than if interference avoidance mechanisms like the reservation of LTE subframes for Wi-Fi only
transmissions are used.

\cite{Almeida2013_EnaLTEWiFi}:Coexistence in TV White Spaces (900MHz), uses almost-blank subframes (ABS) as in LTE Release 10 to accomodate Wi-Fi. Naive allocation and simulation shows Throughput results.

\cite{NokiaVTC2013_perEval}:a performance evaluation of LTE and Wi-Fi coexistence under similar scenarios using a semi-static LTE/Wi-Fi system-level simulator. LTE outperforms Wi-Fi in similar scenarios (including standalone operation of LTE and Wi-Fi). makes Wi-Fi nodes to stay on the LISTEN mode for 96$\%$ of time.
} 
\section{Interference Characterization}
\label{sec:interChara}

\subsection{Interference Characterization Model}
\label{subsec:interChara}
We propose an analytical model to characterize the interference between Wi-Fi and LTE, while considering the Wi-Fi sensing mechanism (clear channel assessment (CCA)) and scheduled and persistent packet transmission at LTE. To illustrate, we focus on a co-channel deployment involving a single W-iFi and a single LTE cell, which involves disseminating the interaction of both technologies in detail and establish a building block to study a complex co-channel deployment of multiple Wi-Fis/LTEs.

In a downlink deployment scenario, a single client and a full buffer (saturated traffic condition) is assumed at each AP under no MIMO\footnotetext[1]{Throughput the paper, LTE home-eNB (HeNB) is also referred as access point (AP) for the purpose of convenience}. Transmit powers are denoted as $P_i, i\in\{w,l\}$ where $w$ and $l$ are indices to denote Wi-Fi and LTE links, respectively. We note that the maximum transmission power of an LTE small cell is comparable to that of the Wi-Fi, and thus is consistent with regulations of unlicensed bands.

The power received from a transmitter $j$ at a receiver $i$ is given by $P_jG_{ij}$ where $G_{ij} \geq 0$ represents a channel gain which is inversely proportional to $d_{ij}^\gamma$ where $d_{ij}$ is the distance between $i$ and $j$ and $\gamma$ is the path loss exponent. $G_{ij}$ may also include antenna gain, cable loss, wall loss, and other factors. Signal-to-Interference-plus-Noise (SINR) on the link $i$ given as
\begin{equation}
\label{sinr}
S_i = \frac{P_iG_{ii}}{P_jG_{ij} + N_i}, \;\; i,j\in\{w,l\},i \neq j
\end{equation}
where $N_i$ is noise power for receiver $i$. Here, in the case of a single Wi-Fi and LTE, if $i$ represents the Wi-Fi link, then $j$ is the LTE link, and vice versa.

The throughput, $R_i, i\in\{w,l\}$, can be represented as a function of $S_i$ as
\begin{equation}
\label{eq:dataRate}
R_i = \alpha_i B \log_2(1 + \beta_iS_i), \;\; i\in\{w,l\},
\end{equation}
where $B$ is a channel bandwidth; $\beta_i$ is a factor  associated with the modulation scheme. For LTE, $\alpha_l$ is a bandwidth efficiency due to factors adjacent channel leakage ratio and practical filter, cyclic prefix, pilot assisted channel estimation, signaling overhead, etc. For Wi-Fi, $\alpha_w$ is the bandwidth efficiency of CSMA/CA, which comes from the Markov chain analysis of CSMA/CA\cite{bianchi2000_802.11DCF} with
\begin{equation}
\label{eq:markov_chain}
\begin{aligned}
\eta_E = \frac{T_E}{\mathrm{E}[S]}, \;
\eta_S = \frac{T_S}{\mathrm{E}[S]}, \;
\eta_C = \frac{T_C}{\mathrm{E}[S]},
\end{aligned}
\end{equation}
where $\mathrm{E}[S]$ is the expected time per Wi-Fi packet transmission; $T_E$, $T_S$, $T_C$ are the average times per $\mathrm{E}[S]$ that the channel is empty due to random backoff, or busy due to the successful transmission or packet collision (in case of multiple Wi-Fis in the CSMA range), respectively. $\alpha_w$ is mainly associated with $\eta_S$.

In our analysis, $\{\alpha_i,\beta_i\}$ is approximated so that - (1) for LTE, $R_l$ matches with throughput achieved under variable channel quality index (CQI), and (2) for Wi-Fi, $R_w$ matches throughput achieved under Biachi's CSMA/CA model.

\subsubsection{Characterization of Wi-Fi Throughput}
Assuming $\lambda_c$ is CCA threshold to detect channel as busy or not, if channel energy at the Wi-Fi node is higher than $\lambda_c$, Wi-Fi would hold back the data transmission, otherwise it transmit at a data rate based on the SINR of the link. Wi-Fi throughput with and without LTE is given as~\\\\
\begingroup
\removelatexerror
\IncMargin{2em}
\begin{algorithm}[H]
\DontPrintSemicolon
\vspace{.3em}
\SetAlgoLined
\SetAlgorithmName{Model}{List of models};
\SetKwInOut{Output}{Output}
\SetKwInOut{Para}{Parameter}

\KwData{$P_w$: Wi-Fi Tx power; $G_w$: channel gain of Wi-Fi link; $P_l$: LTE Tx power; $G_{wl}$: channel gain(LTE AP, Wi-Fi UE); $N_0$: noise power; $E_c$: channel energy at the Wi-Fi (LTE interference + $N_0$).}
\Para{$\lambda_C$: Wi-Fi CCA threshold}
\Output{$R_w$: Wi-Fi throughput}

\eIf{No LTE}{
\vspace{-0.5em}
\[\begin{aligned}
    R_w  = \alpha_w B \log_2\left(1 + \beta_w \frac{P_wG_w}{N_0}\right).
\end{aligned}\]
\vspace{-0.5em}
}
( When LTE is present){\eIf{$E_c >$ $\lambda_C$}{
No Wi-Fi transmission with $R_w = 0$
}{
\vspace{-0.5em}
\[\begin{aligned}
	R_w  = \alpha_w B \log_2\left(1 + \beta_w \frac{P_wG_w}{P_lG_{wl} + N_0}\right).
\end{aligned}\]
\vspace{-1em}
}}
\caption{Wi-Fi Throughput Characterization}
\end{algorithm}
\DecMargin{2em}
\endgroup

\subsubsection{Characterization of LTE Throughput}~
Due to CSMA/CA, Wi-Fi is active for an average $\eta_S$ fraction of time (Eq. \eqref{eq:markov_chain}). Assuming that LTE can instantaneously update its transmission rate based on the Wi-Fi interference, its throughput can be modeled as follows- ~\\\\
\begingroup
\removelatexerror
\IncMargin{2em}
\begin{algorithm}[H]
\DontPrintSemicolon
\vspace{.3em}
\SetAlgoLined
\SetAlgorithmName{Model}{List of models};
\SetKwInOut{Output}{Output}
\SetKwInOut{Para}{Parameter}

\KwData{$P_l$: LTE Tx power; $G_l$: channel gain of LTE link; $P_w$: Wi-Fi Tx power; $G_{lw}$: channel gain(Wi-Fi AP,LTE UE); $N_0$: noise power; $E_c$: channel energy at Wi-Fi (LTE interference + $N_0$);}
\Para{$\lambda_C$: Wi-Fi CCA threshold}
\Output{$R_l$: LTE throughput}

\eIf{No Wi-Fi}{
\vspace{-0.5em}
\[\begin{aligned}
    R_{l_{\mbox{noW}}}  &= \alpha_l B \log_2\left(1 + \beta_l \frac{P_lG_l}{N_0}\right).
\end{aligned}\]
\vspace{-0.5em}
}
( When Wi-Fi is present){\eIf{$E_c >$ $\lambda_C$}{
No Wi-Fi transmission/interference
\[\begin{aligned}
    R_l &= R_{l_{\mbox{noW}}}.
\end{aligned}\]
\vspace{-1.2em}
}{
\vspace{-0.5em}
\[\begin{aligned}
    R_l  &= \alpha_l B \log_2\left(1 + \beta_l \frac{P_lG_l}{P_lG_{lw} + N_0}\right).
\end{aligned}\]
Using \eqref{eq:markov_chain} and $\eta_C = 0$ (a single Wi-Fi)
\[\begin{aligned}
    R_l &= \eta_E R_{l_{\mbox{noW}}} +  \eta_S R_l
\end{aligned}\]
\vspace{-1.2em}
}}
\caption{LTE Throughput Characterization}
\end{algorithm}
\DecMargin{2em}
\endgroup

\begin{figure}[t]
\begin{center}
\includegraphics[height=1.5in,width=3in]{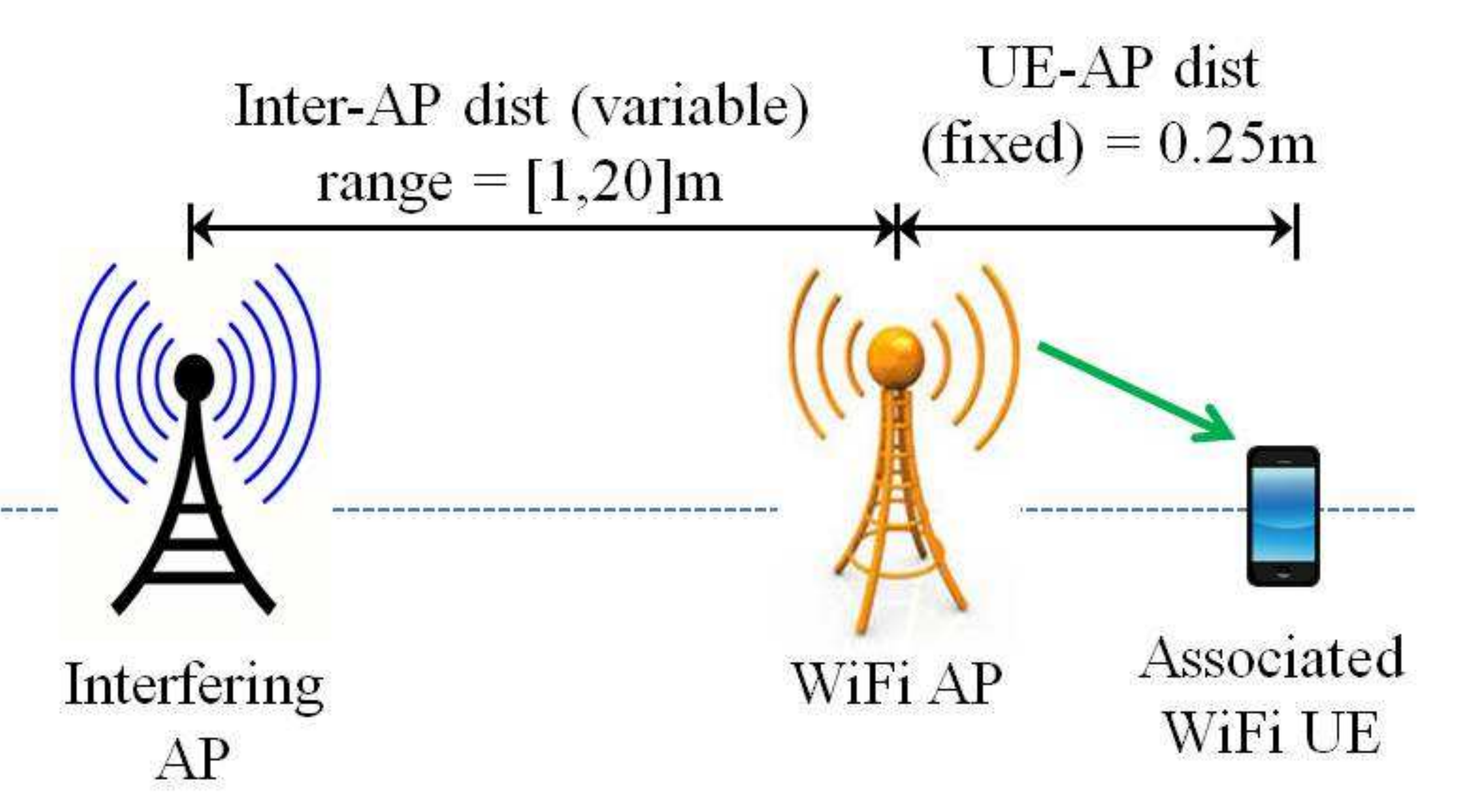}
\caption{Experimental scenario to evaluate the throughput performance of Wi-Fi $w_1$ in the presence of interference (LTE/other Wi-Fi/white noise) when both $w_1$ and interference operated on the same channel in 2.4 GHz}
\label{fig:scene1_interonWiFi}
\end{center}
\vspace{-0.3cm}
\end{figure}

\begin{figure}[t]
\begin{center}
\includegraphics[height=2in,width=2.75in]{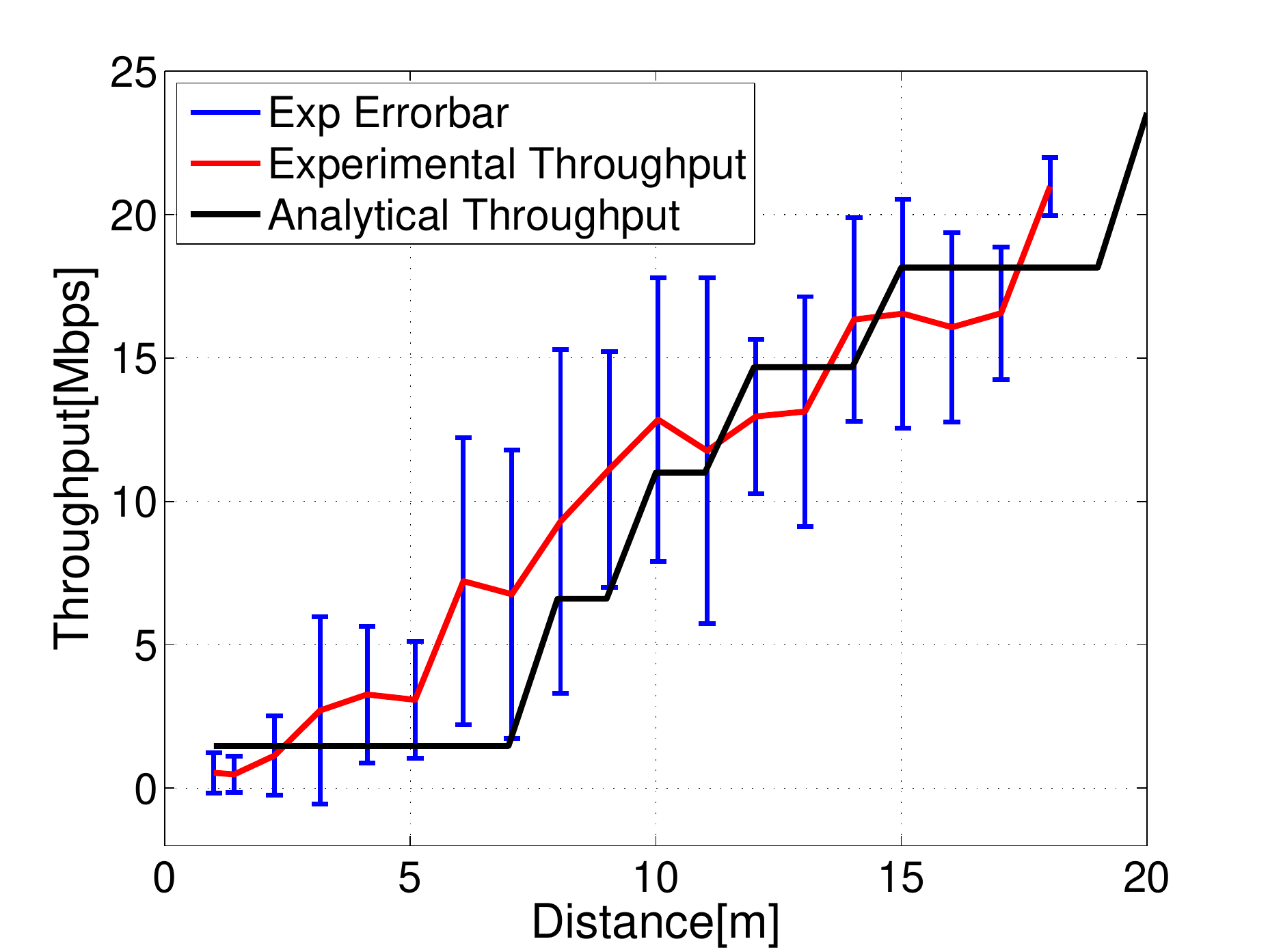}
\caption{Comparative results analytical model and experiments to show the effect of LTE on the throughput of Wi-Fi 802.11g  when distance between LTE eNB and Wi-Fi link is varied.}
\label{fig:compAnaModelvsExp_11g}
\end{center}
\end{figure}

\begin{figure}[t]
\begin{center}
\includegraphics[height=2in,width=2.75in]{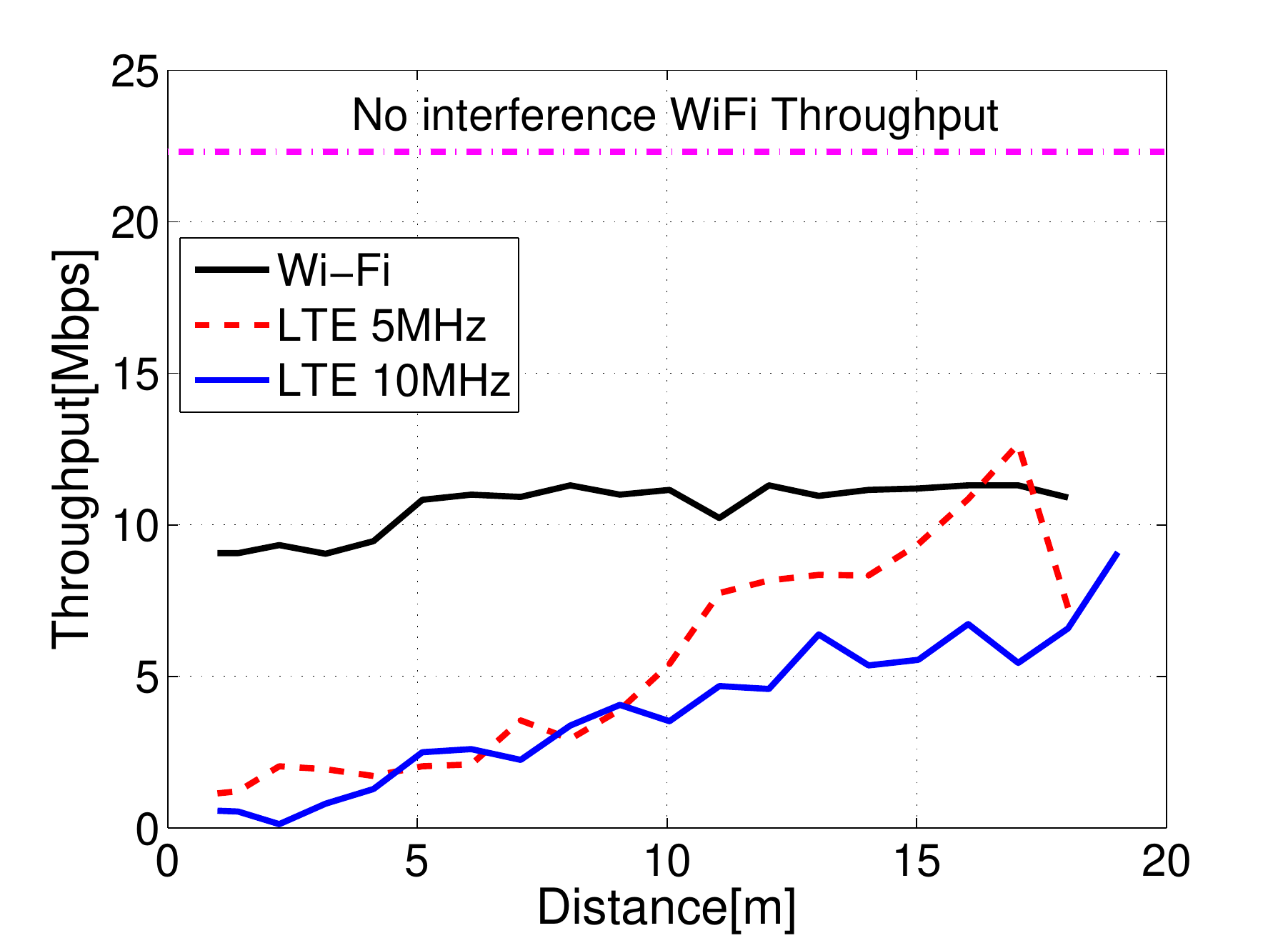}
\caption{Comparative results analytical model and experiments to show the effect of LTE on the throughput of Wi-Fi 802.11g  when distance between LTE HeNB (AP) and Wi-Fi link is varied.}
\label{fig:distvsThru_11g_MME}
\end{center}
\end{figure}

\begin{table}[t]
\caption{Network parameters of Wi-Fi/LTE deployment}
\centering
\begin{tabular}{| l | l | l | l |}
\hline
\textbf{Parameter} & \textbf{Value} & \textbf{Parameter} & \textbf{Value}\\
\hline
Scenario & Downlink & Tx power & 20 dBm\\
\hline
Spectrum band & 2.4 GHz & Channel bandwidth & 20 MHz\\
\hline
Traffic model & \multicolumn{3}{|l|}{Full buffer via saturated UDP flows}\\
\hline
AP antenna height & 10 m & User antenna height & 1 m\\
\hline
Path loss model & \multicolumn{3}{|l|}{36.7$\log_{10}$(d[m]) + 22.7 + 26$\log_{10}$(frq [GHz])}\\
\hline
Noise Floor & \multicolumn{3}{|l|}{-101 dBm, (-174 cBm thermal noise/Hz)}\\
\hline
Channel & \multicolumn{3}{|l|}{No shadow/Rayleigh fading}\\
\hline
Wi-Fi & \multicolumn{3}{|l|}{802.11n: SISO}\\
\hline
LTE & \multicolumn{3}{|l|}{FDD, Tx mode-1 (SISO)}\\
\hline
\end{tabular}
\label{tab:net_para}
\end{table}

\subsection{Experimental Validation}
In this section, we experimentally validate proposed interference characterization models using experiments involving the ORBIT testbed and USRP radio platforms available at WINLAB\cite{Ray2005_ORBITOverview,usrpB210}. An 802.11g Wi-Fi link is set up  using Atheros AR928X wireless network adapters\cite{ath9k} and an AP implementation with \textit{hostapd}\cite{hostapd}. For LTE, we use \textit{OpenAirInterface}, an open-source software implementation, which is fully compliant with 3GPP LTE standard (release 8.6) and set in transmission mode 1 (SISO)\cite{Nikaein2014_OAI}. Currently, \textit{OpenAirInterface} is in the development mode for USRP based platforms with limited working LTE operation parameters.

In our experiment, depicted as the scenario shown in figure~\ref{fig:scene1_interonWiFi}, we study the effect of interference on the Wi-Fi link $w_1$. For link $w_1$, the distance between the AP and client is fixed at $0.25$ m (very close so that the maximum throughput is guaranteed when interference is present. Experimentally, we observe maximum throughput as $22.2$ Mbps). The distance between the interfering AP and Wi-Fi AP is varied in the range of 1 to 20 m. The throughput of $w_1$ is evaluated under three sources of interference - LTE and Wi-Fi, when both $w_1$ and the interference AP is operated on the same channel in the 2.4 GHz spectrum band. These experiments are carried in the 20 m-by-20 m ORBIT room in WINLAB, which has an indoor Line-of-Sight (LoS) environment. For each source of interference, Wi-Fi throughput is averaged over 15 sets of experiments with variable source locations and trajectories between interference and $w_1$.

In the first experiment, we perform a comparison study to evaluate the effect of LTE interference on $w_1$, observed by experiments and computed by interference characterization model. In this case, LTE signal is lightly loaded on 5 MHz of bandwidth mainly consist of control signals. Thus, the impact of such LTE signal over the Wi-Fi band is equivalent to the low power LTE transmission. Thus, we incorporate these LTE parameters in our analytical model. As shown in figure~\ref{fig:compAnaModelvsExp_11g}, we observe that both experimental and analytical values match the trend very closely, though with some discrepancies. These discrepancies are mainly due to the fixed indoor experiment environment and lack of a large number of experimental data sets. Additionally, we note that even with the LTE control signal (without any scheduled LTE data transmission), performance of Wi-Fi gets impacted drastically.

In the next set of experiments, we study the throughput of a single Wi-Fi link in the presence of different sources of interference - (1) Wi-Fi, (2) LTE operating at 5 MHz, and (3) LTE operating at 10 MHz, evaluating each case individually. For this part, full-band occupied LTE is considered with the maximum power transmission of $100$ mW. As shown in figure~\ref{fig:distvsThru_11g_MME}, when the Wi-Fi link operates in the presence of other Wi-Fi links, they share channel according to the CSMA/CA protocol and throughput is reduced approximately by half. In the both the cases of LTE operating at 5 and 10 MHz, due to lack of coordination, Wi-Fi throughput gets impacted by maximum upto $90\%$ compared to no interference Wi-Fi throughput and $20-80\%$ compared to Wi-Fi thorughput in the presence of other Wi-Fi link. 
These results indicate significant inter-system interference in the baseline case without any coordination between systems. 

\subsection{Motivational Example}
\label{subsec:motiEx}
\begin{figure}[t]
\begin{center}
\includegraphics[height=1.5in,width=3.5in]{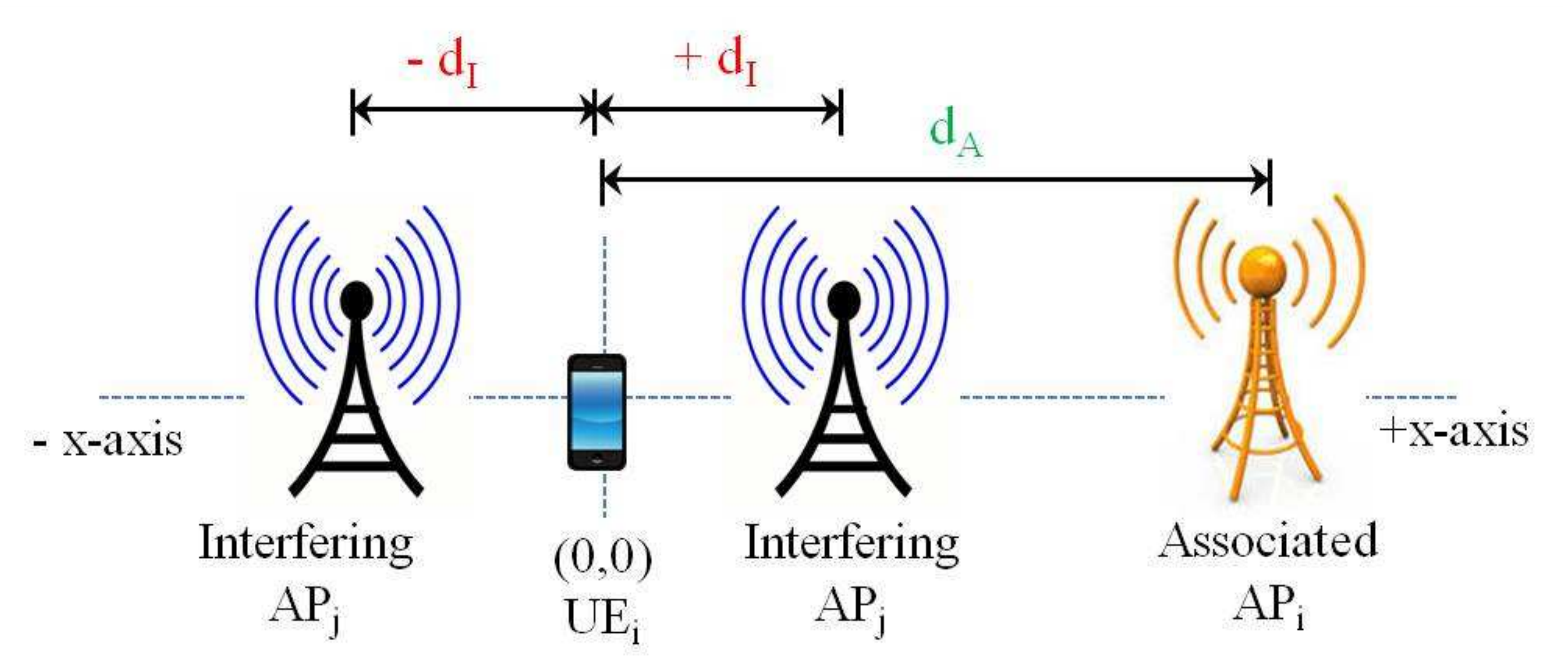}
\caption{Experimental scenario to evaluate the throughput performance of Wi-Fi $w_1$ in the presence of interference (LTE/other Wi-Fi/white noise) when both $w_1$ and interference operated on the same channel in 2.4 GHz}
\label{fig:scene2_varIntAPUEdist}
\end{center}
\end{figure}

We extend our interference model to complex scenarios involving co-channel deployment of a single link Wi-Fi and LTE for the detailed performance evaluation. As shown in figure~\ref{fig:scene2_varIntAPUEdist}, UE$_i$, associated AP$_i$ and interfering AP$_j$, $i,j \in \{w,l\}, i \neq j$, are deployed in a horizontal alignment. The distance, $d_A$, between UE$_i$ and AP$_i$ is varied between $0$ and $100$ m. At each value of $d_A$, the distance between UE$_i$ and AP$_j$ is varied in the range of $-100$ to $100$ m. Assuming UE$_i$ is located at the origin $(0,0)$, if AP$_j$ is located on the negative X-axis then the distance is denoted as $-d_I$, otherwise as $+d_I$, where $d_I$ is an Euclidean norm $\|\text{UE}_i,\text{AP}_j\|$. In the shared band operation of Wi-Fi and LTE, due to the CCA sensing mechanism at the Wi-Fi node, the distance between Wi-Fi and LTE APs (under no shadow fading effect in this study) decides the transmission or shutting off of Wi-Fi. Thus, the above distance convention is adopted to embed the effect of distance between AP$_i$ and AP$_j$. Simulation parameters for this set of simulations are given in Table~\ref{tab:net_para}.

\begin{figure}[t]
\centering
\subfigure[A heat map of Wi-Fi throughput (Mbps)]{
	\includegraphics[height=1.75in,width=2.3in]{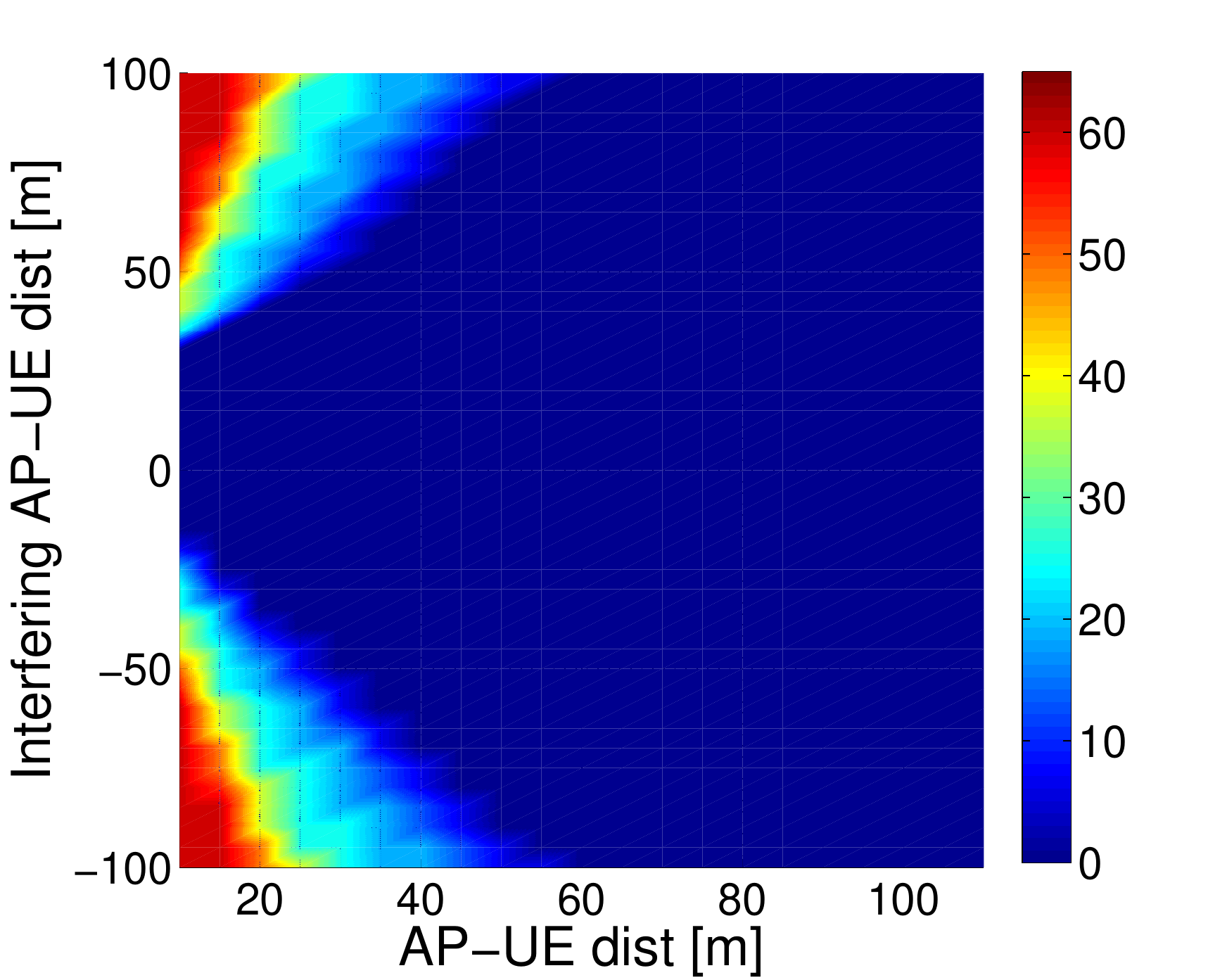}
    \label{fig:wthru_lteInter_3D}
}
\subfigure[Wi-Fi performance sections- High SINR: non-zero throughput, Low SINR: SINR below minimum SINR requirement, CCA$\_$busy: shutting off of Wi-Fi due to channel is sensed as busy]{
	\includegraphics[height=1.75in,width=2.3in]{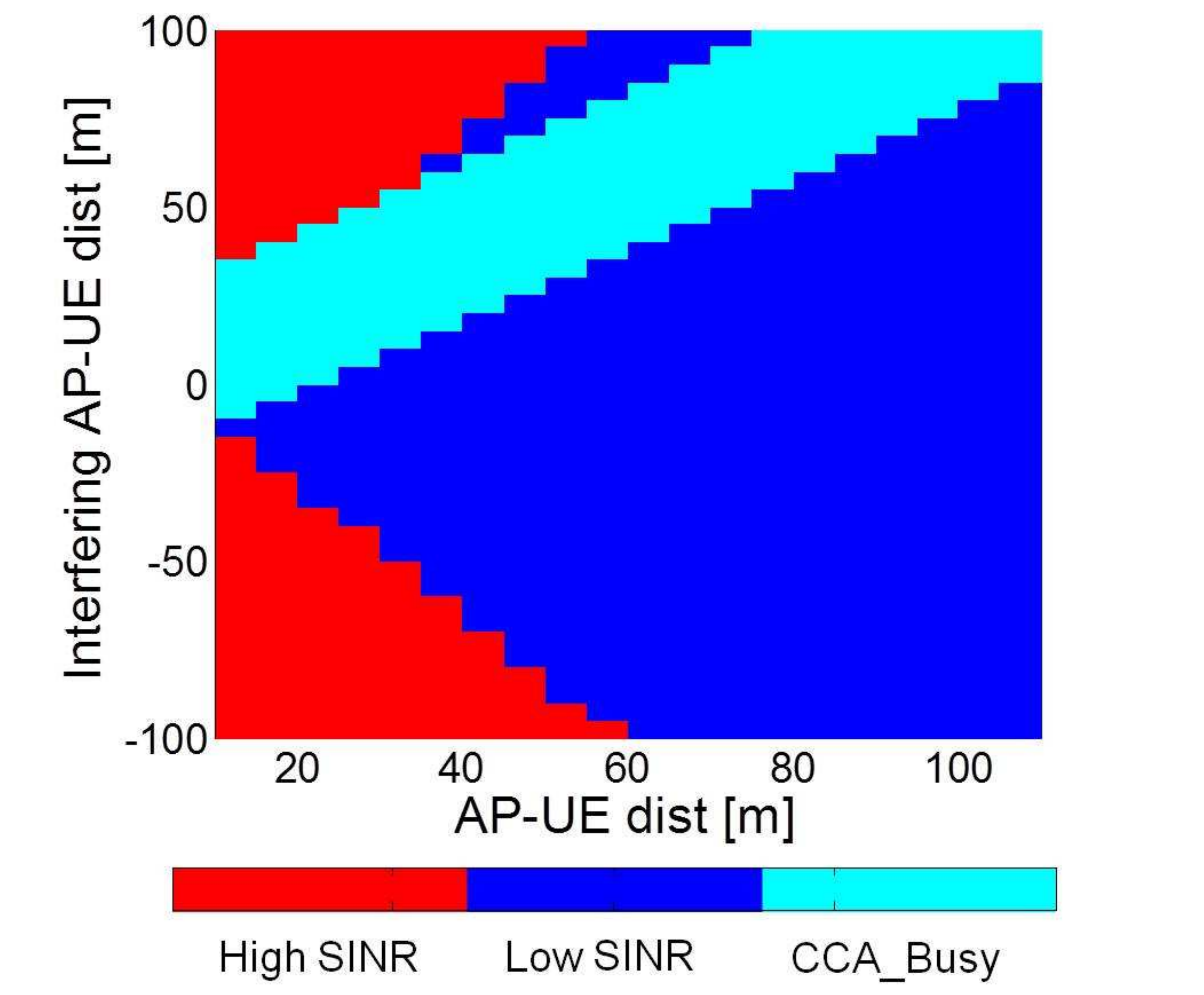}
    \label{fig:w_perfSections}
}
\caption[Optional caption for list of figures]{Wi-Fi performance as a function of distance(Wi-Fi AP, associated Wi-Fi UE) $d_A$ and distance(Interfering LTE AP, Wi-Fi UE) $d_I$  
}
\label{fig:wifi_noOpt}
\end{figure}

Figure~\ref{fig:wifi_noOpt} shows the Wi-Fi performance in the presence of LTE interference. As shown in figure~\ref{fig:wthru_lteInter_3D}, the Wi-Fi throughput is drastically deteriorated in the co-channel LTE operation, leading to zero throughput for $80\%$ of the cases and an average $91\%$ of throughput degradation compared to standalone operation of Wi-Fi. Such degradation is explained by figure~\ref{fig:w_perfSections}. Region \textit{CCA$\_$busy} shows the shutting off of the Wi-Fi AP due to the CCA mechanism, where high energy is sensed in the Wi-Fi band. This region corresponds to cases when Wi-Fi and LTE APs are within $\sim 20$m of each other. In the \textit{low SINR} region, the Wi-Fi link does not satisfy the minimum SINR requirement for data transmission, thus the Wi-Fi throughput is zero. \textit{High SINR} depicts the data transmission region that satisfies SINR and CCA requirements and throughput is varied based on variable data rate/SINR.

\begin{figure}[t]
\centering
\subfigure[A heat map of LTE throughput (Mbps)]{
	\includegraphics[height=1.75in,width=2.3in]{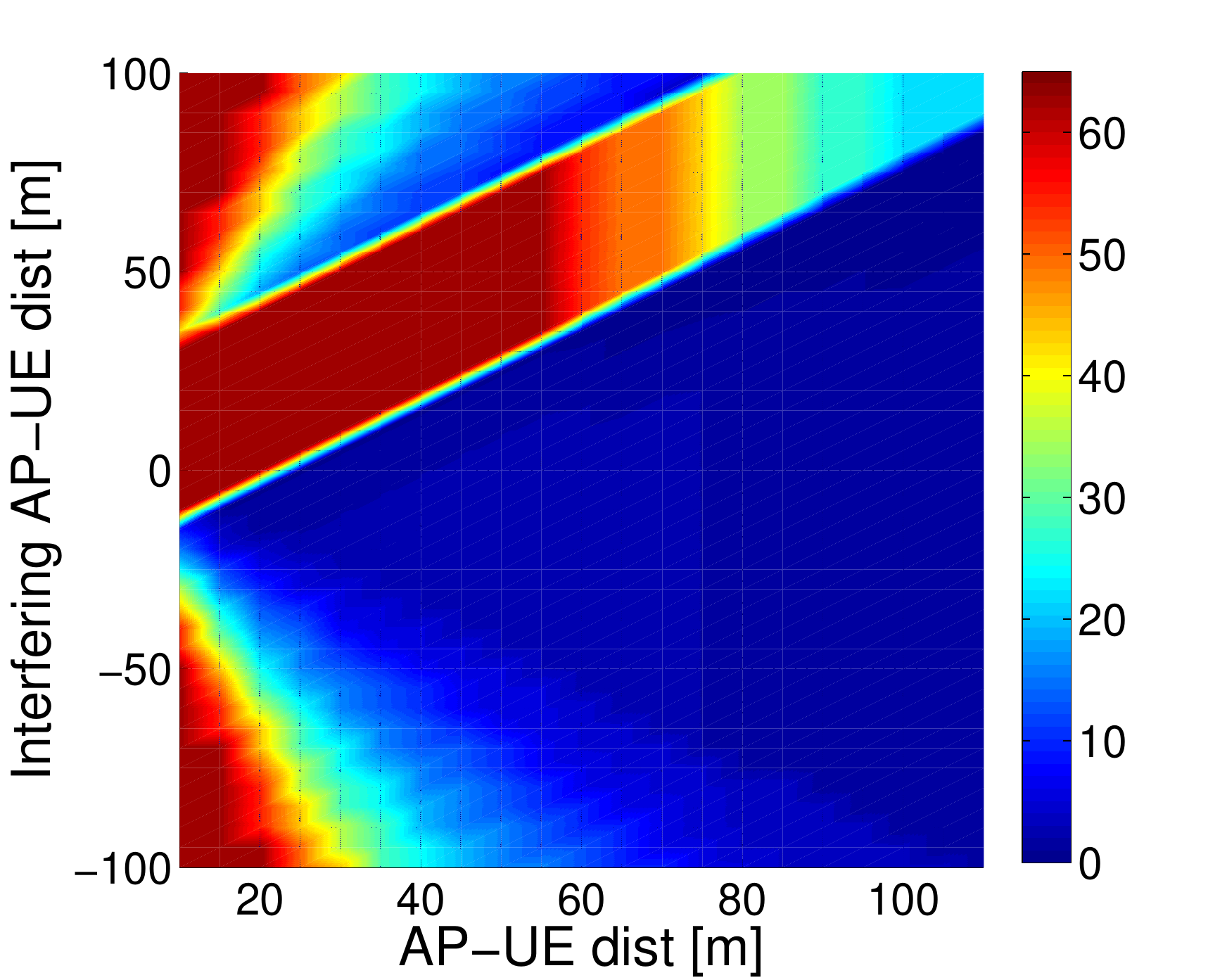}
    \label{fig:Lthru_lteInter_3D}
}
\subfigure[LTE performance sections- High SINR: non-zero throughput, Low SINR: SINR below minimum SINR requirement, CCA$\_$busy: shutting off of Wi-Fi due to channel is sensed as busy]{
	\includegraphics[height=1.75in,width=2.3in]{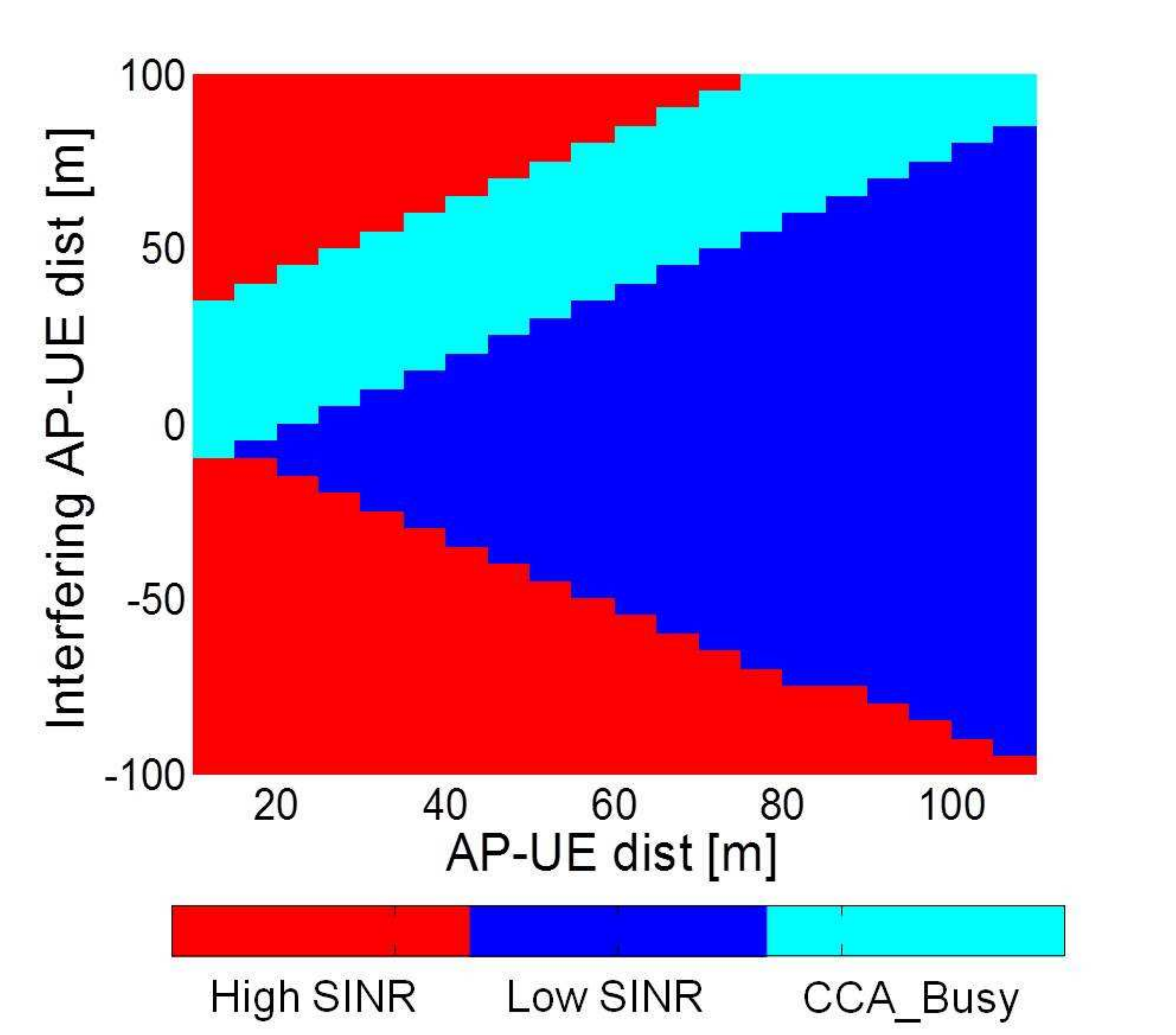}
    \label{fig:L_perfSections}
}
\caption[Optional caption for list of figures]{LTE performance as a function of distance(LTE AP, associated LTE UE) $d_A$ and distance(Interfering Wi-Fi AP, LTE UE) $d_I$  
}
\label{fig:lte_noOpt}
\end{figure}

On the other hand, figure~\ref{fig:lte_noOpt} depicts the LTE throughput in the presence of Wi-Fi interference.  LTE throughput is observed to be zero in the \textit{low SINR} regions, which is $45\%$ of the overall area and the average throughput degradation is $65\%$ compared to the standalone LTE operation. Under identical network parameters, overall performance degradation for LTE is much lower compared to that of Wi-Fi in the previous example. The reasoning for such a behavior discrepancy is explained with respect to figure~\ref{fig:L_perfSections} and the Wi-Fi CCA mechanism. In the \textit{CCA$\_$busy} region, Wi-Fi operation is shut off and LTE operates as if no Wi-Fi is present. In both LTE and the previous Wi-Fi examples, \textit{low SINR} represents the hidden node problem where two APs do not detect each other's presence and data transmission at an UE suffers greatly.

\begin{figure*}
\begin{center}
\includegraphics[height=2.6in,width=6.2in]{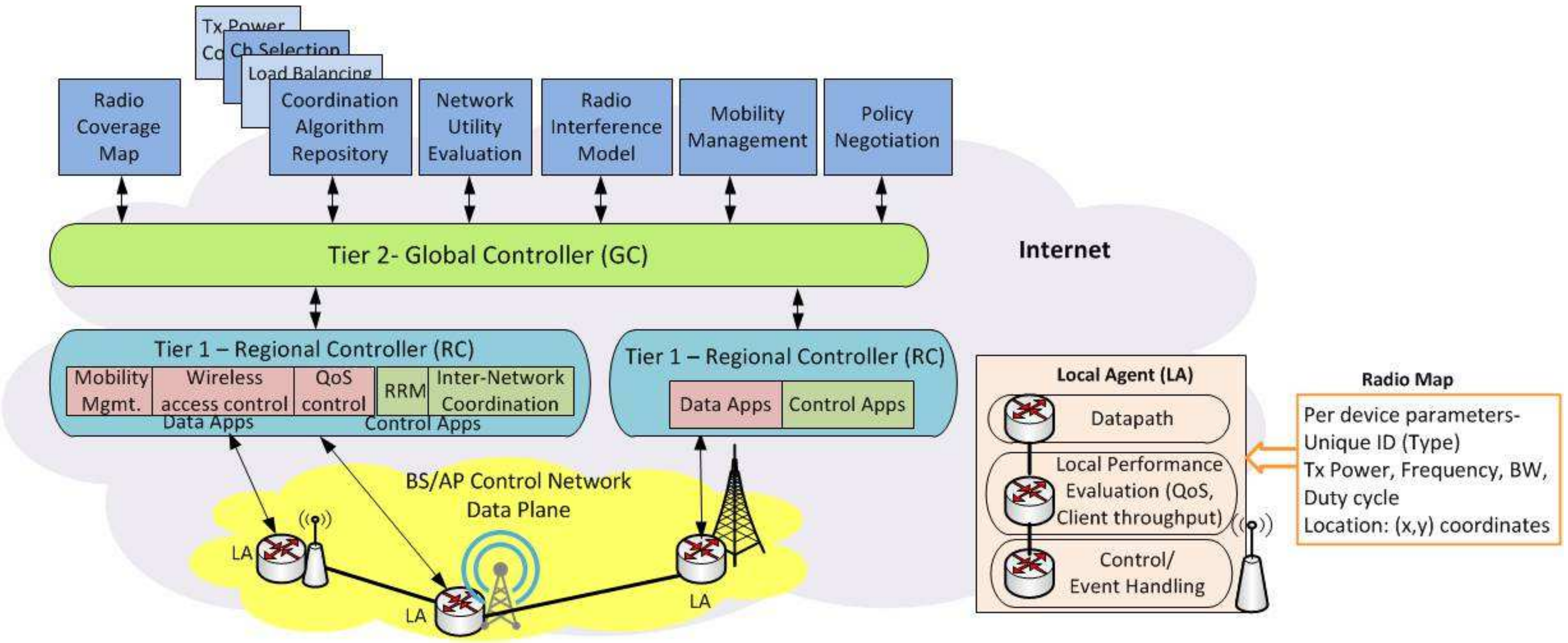}
\caption{SDN based achitecture for inter-network cooperation on radio resource management}
\label{fig:SDN_arch}
\end{center}
\end{figure*}

\section{System Architecture}
\label{sec:arch}

In this section, we describe an architecture for coordinating between multiple heterogeneous networks to improve spectrum utilization and facilitate co-existence\cite{Raychaudhuri2014_NASCOR}. Figure \ref{fig:SDN_arch} shows the proposed system, which is built on the principles of a Software Defined Networking (SDN) architecture to support logically-centralized dynamic spectrum management involving multiple autonomous networks. The basic design goal of this architecture is to support the seamless communication and information dissemination required for coordination of heterogeneous networks. The system consists of two-tiered controllers: the Global Controller (GC) and Regional Controllers (RC), which are mainly responsible for the control plane of the architecture. The GC, owned by any neutral/authorized organization, is the main decision making entity, which acquires and processes network state information and controls the flow of information between RCs and databases based on authentication and other regulatory policies. Decisions at the GC are based on different network modules, such as radio coverage maps, coordination algorithms, policy and network evaluation matrices. The RCs are limited to network management of specific geographic regions and the GC ensures that RCs have acquired local visibility needed for radio resource allocation at wireless devices.
A Local Agent (LA) is a local controller, co-located with an access point or base-station. It receives frequent spectrum usage updates from wireless clients, such as device location, frequency band, duty cycle, power level, and data rate. The signaling between RC and LAs are event-driven, which occurs in scenarios like the non-fulfillment of quality-of-service (QoS) requirements at wireless devices, request-for-update from an RC and radio access parameter updates from an RC. The key feature of this architecture is that the frequency of signaling between the different network entities is less in higher tiers compared to lower tiers. RCs only control the regional messages and only wide-area network level signalling protocols are handled at the higher level, GC. Furthermore, this architecture allows adaptive coordination algorithms based on the geographic area and change in wireless device density and traffic patterns. We use this architecture to exchange control messages required for the optimization model, as described in \S \ref{sec:optimize}.

\section{System Model}
\label{sec:sysModel}
As seen in the previous section, when two (or more) APs of different Wi-Fi and LTE networks are deployed in the same spectrum band, APs can cause severe interference to one another. In order to alleviate inter-network interference, we propose joint coordination based on (1) power, and (2) time division channel access optimization. We assume that both LTE and Wi-Fi share a single spectrum channel and operate on the same amount of bandwidth. We also note that clients associated to one AP cannot join other Wi-Fi or LTE APs. This is a typical scenario when multiple autonomous operators deploy APs in the shared band. With the help of the proposed SDN architecture, power level and time division channel access parameters are forwarded to each network based on the throughput requirement at each UE. To the best of our knowledge, such an optimization framework has not yet received much attention for the coordination between Wi-Fi and LTE networks.

We consider a system with $N$ Wi-Fi and $M$ LTE networks. $\mathcal{W}$ and $\mathcal{L}$ denote the sets of Wi-Fi and LTE links, respectively. We maintain all assumptions, definitions and notations as described in Section \ref{subsec:interChara}. For notational simplicity, we redefine $R_i = \alpha_i B \log_2(1 + \beta_i S_i), i\in\{\mathcal{W,L}\}$ as $R_i = \alpha_i \log_2(1 + \beta_i S_i)$, where constant parameter $B$ is absorbed with $\alpha_i$. Additional notation are summarized in Table \ref{tab:optNot}.
\begin{table}[t]
\caption{Definition of notations}
\centering
\begin{tabular}{| l | l |}
\hline
\textbf{Notation} & \textbf{Definition}\\
\hline
$w, l$ & indices for Wi-Fi and LTE network, respectively\\
\hline
$\mathcal{W}$ & the set of Wi-Fi links\\
\hline
$\mathcal{L}$ & the set of LTE links \\
\hline
$P_i$ & Transmission power of $i$-th AP, where $i\in\{\mathcal{W,L}\}$\\
\hline
$G_{ij}$ & Channel gain between nodes $i$ and $j$ \\
\hline
$R_i$ & Throughput at $i$-th link, where $i\in\{\mathcal{W,L}\}$\\
\hline
$S_i$ & SINR at $i$-th link, where $i\in\{\mathcal{W,L}\}$\\
\hline
$B$ & Channel Bandwidth\\
\hline
$N_0$ & Noise level \\
\hline
$\alpha_i,\beta_i$ & Efficiency parameters of system $i\in\{\mathcal{W,L}\}$\\
\hline
$M^a_i$ & Set of Wi-Fi APs in the CSMA range of AP $i\in\{\mathcal{W}\}$ \\
\hline
$M^b_i$ & Set of Wi-Fi APs in the interference range of AP $i\in\{\mathcal{W}\}$\\
\hline
$\zeta$ & Hidden node interference parameter\\
\hline
$\eta$ & Fraction of channel access time for network $i,i\in\{w,l\}$ when\\
& $j,j\in\{w,l\},j \neq i$, access channel for $1-\eta$ fraction of time\\
\hline
\end{tabular}
\label{tab:optNot}
\end{table}

In order to account for the co-channel deployment of multiple Wi-Fi networks, we assume that time is shared equally when multiple Wi-Fi APs are within CSMA range due to the Wi-Fi MAC layer. We denote the set of Wi-Fi APs within the CSMA range of AP$_i, i\in\{\mathcal{W}\}$ as $M^a_i$ and those outside of carrier sense but within interference range as $M^b_i$. When AP$_i$ shares the channel with $|M^a_i|$ other APs, its share of the channel access time get reduced to approximately $1/(1 + |M^a_i|)$. Furthermore, $M^b_i$ signifies a set of potential hidden nodes for AP$_i, \forall i$. To capture the effect of hidden node interference from APs in the interference range, parameter $\zeta$ is introduced which lowers the channel access time and thus, the throughput. Average reduction in channel access time at AP$_i$ is $1/(1 + \zeta|M^b_i|)$ where $\zeta$ falls in the range $[0.2,0.6]$\cite{baid2012_netCoop}. Therefore, the effective Wi-Fi throughput can be written as

\begin{equation}
\label{eq:w_thru}
\begin{aligned}
R_i = a_i b_i \alpha_w \log_2(1 + \beta_w S_i), \;\; i\in \mathcal{W},\\
\text{with } a_i = \frac{1}{1 + |M^a_i|} \text{ and } b_i = \frac{1}{1 + \zeta|M^b_i|}.
\end{aligned}
\end{equation}
SINR of Wi-Fi link, $i, i\in\mathcal{W}$, in the presence of LTE and no LTE is described as
\begin{equation}
\label{eq:w_sinr}
S_i = \left\{ \begin{aligned}
&\frac{P_iG_{ii}}{N_0}, && \text{if no LTE};\\
&\frac{P_iG_{ii}}{\sum_{j\in \mathcal{L}}P_jG_{ij} + N_0}, && \text{if LTE},
\end{aligned}
\right.
\vspace{0.05cm}
\end{equation}
where the term $\sum_{j\in L}P_jG_{ij}$ is the interference from all LTE networks at a Wi-Fi link $i$.

The throughput definition of the LTE link $i, i\in\mathcal{L}$ remains the same as in Section \ref{subsec:interChara}:
\[
R_i = \alpha_l \log_2(1 + \beta_l S_i), \;\; i\in\mathcal{L}.
\]
The SINR of the LTE link, $i, \forall i$, in the presence of Wi-Fi and no Wi-Fi is described as
\begin{equation}
\label{eq:l_sinr}
S_i = \left\{ \begin{aligned}
&\frac{P_iG_{ii}}{\sum_{j\in \mathcal{L}, j \neq i}P_jG_{ij} + N_0}, \; \text{if no Wi-Fi};\\
&\frac{P_iG_{ii}}{\sum_{j\in \mathcal{L}, j \neq i}P_jG_{ij} + \sum_{k\in \mathcal{W}}a_kP_kG_{ik} + N_0},\; \text{if Wi-Fi},
\end{aligned}
\right.
\vspace{0.05cm}
\end{equation}
where terms $\sum_{j\in\mathcal{L}, j \neq i}P_jG_{ij}$ and $\sum_{k\in\mathcal{W}}a_kP_kG_{ik}$ signifies the interference contribution from other LTE links and Wi-Fi links, (assuming all links in $\mathcal{W}$ are active). For the $k$-th Wi-Fi link, $\forall k$, the interference is reduced by a factor $a_k$ to capture the fact that the $k$-th Wi-Fi is active approximately for only $a_k$ fraction of time due to the CSMA/CA protocol at Wi-Fi.

For a given model, inter-network coordination is employed to assure a minimum throughput requirement, thus the guaranteed availability of the requested service at each UE. For this purpose, we have implemented our optimization in two stages as described in following subsections.
\eat{Using SDN architecture, transmission power of Wi-Fi and LTE can be dynamically changed based on the global visibility of network with respect to geographical placement of Wi-Fi/LTE nodes, interference pattern at nodes, throughput requirement at each UE/application, etc.}
\section{Coordination via Joint Optimization}
\label{sec:optimize}

\subsection{Joint Power Control Optimization}
Here, the objective is to optimize the set of transmission power $P_i,i\in\{\mathcal{W,L}\}$ at Wi-Fi and LTE APs, which maximizes the aggregated Wi-Fi+LTE throughput. Conventionally, LTE supports the power control in the cellular network. By default, commercially available Wi-Fi APs/routers are set to maximum level\cite{iw}. But adaptive power selection capability is incorporated in available 802.11a/g/n Wi-Fi drivers, even though it is not invoked very often. Under the SDN architecture, transmission power level can be made programmable to control the influence of interference from any AP at neighboring radio devices based on the spectrum parameters\cite{Gudipati2013_SoftRAN}.

For the maximization of aggregated throughput, we propose a geometric programming (GP) based power control\cite{Chiang2007_pwrCtr}. For the problem formulation, throughput, given by Eq.~\ref{eq:dataRate}, can approximated as
\begin{equation}
\label{eq:dataRate_1}
\begin{aligned}
R_i = \alpha_i \log_2(\beta_iS_i), \;\;i\in\{\mathcal{W,L}\}.
\end{aligned}
\end{equation}
This equation is valid when $\beta_iS_i$ is much higher than 1. In our case, this approximation is reasonable considering minimum SINR requirements for data transmission at both Wi-Fi and LTE. 
The aggregate throughput of the WiFi+LTE network is
\begin{equation}
\label{eq:dataRate_sys}
\begin{aligned}
\mathbb{R} &= \sum_{i\in\mathcal{W}} a_i b_i \alpha_w \log_2(\beta_w S_i) + \sum_{j\in\mathcal{L}} \alpha_l \log_2(\beta_l S_j)\\
&= \log_2\left[\left(\prod_{i\in\mathcal{W}} (\beta_w S_i)^{a_i b_i \alpha_w} \right) \left(\prod_{j\in\mathcal{L}} (\beta_l S_i)^{\alpha_l} \right)\right].
\end{aligned}
\end{equation}

In the coordinated framework, it is assumed that WiFi parameters $a_i$ and $b_i$ are updated periodically. Thus, these are considered as constant parameters in the formulation. Also, $\alpha_i, \beta_i, i\in\{w,l\}$ are constant in the network. Therefore, aggregate throughput maximization is equivalent to maximization of a product of SINR at both WiFi and LTE links. Power control optimization formulation is given by:
\begin{equation}
\begin{aligned}
& {\text{maximize}}
& & \left(\prod_{i\in\mathcal{W}} (\beta_w S_i)^{a_i b_i \alpha_w} \right) \left(\prod_{j\in\mathcal{L}} (\beta_l S_i)^{\alpha_l} \right)\\
& \text{subject to}
& & R_i \geq R_{i,\min}, \;\; i\in\mathcal{W},\\
&&& R_i \geq R_{i,\min}, \;\; i\in\mathcal{L},\\
&&& {\sum_{k\in M^b_i } P_kG_{ik}+\sum_{j\in\mathcal{L}} P_jG_{ij}} + N_0 < \lambda_c, \; i\in W,\\
&&& 0 < P_i \leq P_{\max}, \;\; i\in\mathcal{W},\\
&&& 0 < P_i \leq P_{\max}, \;\; i\in\mathcal{L}.\\
\end{aligned}
\end{equation}

Here, the first and second constraints are equivalent to $S_i \geq S_{i,min}, \forall i$ which ensures that SINR at each link achieves a minimum SINR requirement, thus leading to non-zero throughput at the UE. The third constraint assures that channel energy at a WiFi (LTE interference + interference from WiFis in the interference zone + noise power) is below the clear channel assessment threshold $\lambda_c$, thus WiFi is not shut off. The fourth and fifth constraints follow the transmission power limits at each link. Unlike past power control optimization formulations for cellular networks, WiFi-LTE coexistence requires to meet the SINR requirement at a WiFi UE and, additionally, CCA threshold at a WiFi AP. 

For multiple Wi-Fi and LTE links, to ensure the feasibility of the problem where all constrains are not satisfied, notably for WiFi links, we relax the minimum data requirement constraint for LTE links. In our case, we reduce the minimum data requirement to zero. This is equivalent to shutting off certain LTE links which cause undue interference to neighboring WiFi devices.

\subsection{Joint Time Division Channel Access Optimization}
The relaxation of minimum throughput constraint in the joint power control optimization leads to throughput deprivation at some LTE links. Thus, joint power control is not sufficient when system demands to have non-zero throughput at each UE. In such cases, we propose a time division channel access optimization framework where network of each RAT take turns to access the channel. Assuming network $i, i\in \{w,l\}$ access the channel for $\eta, eta \in [0,1],$ fraction of time, network $j, j\in\{w,l\},j\neq i,$ holds back the transmission and thus no interference occurs at $i$ from $j$. For remaining $1-\eta$ fraction of time, $j$ access the channel without any interference from $i$. This proposed approach can be seen as a subset of power assignment problem, where power levels at APs of network $i, i\in\{w,l\},$ is set to zero in their respective time slots. The implementation of the protocol is out of scope of this paper.

In this approach, our objective is to optimize $\eta$, the time division of channel access, such that it maximizes the minimum throughput across both WiFi and LTE networks. We propose the optimization in two steps -

\begin{figure*}[t]
\begin{center}
\subfigure[A heat map of WiFi throughput when joint power Coordination (Mbps)]{
	\includegraphics[height=1.75in,width=2.3in]{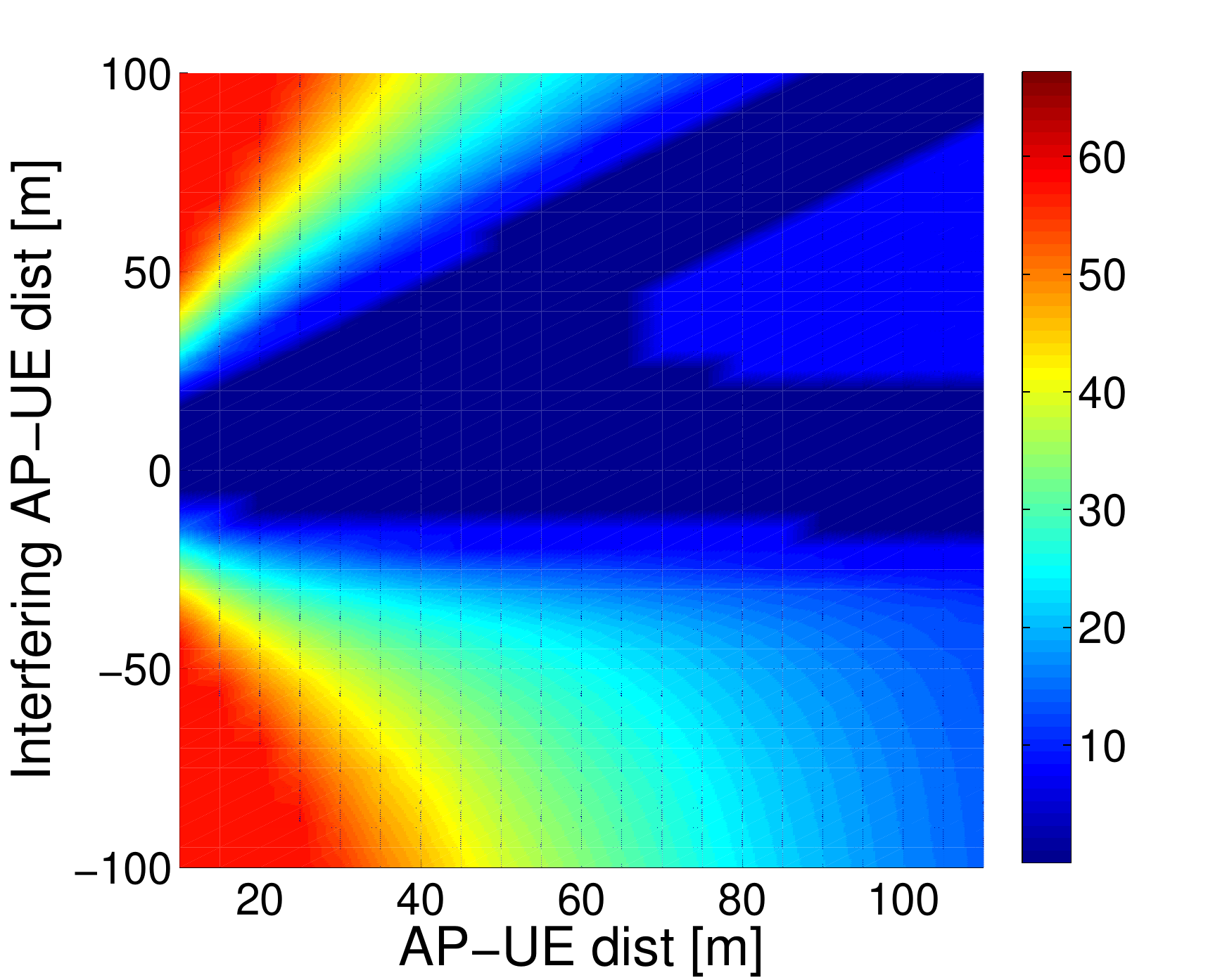}
    \label{fig:W_wiL_pwrOpt_3d}
}
\subfigure[Feasibility region of joint power Coordination]{
	\includegraphics[height=1.75in,width=2.1in]{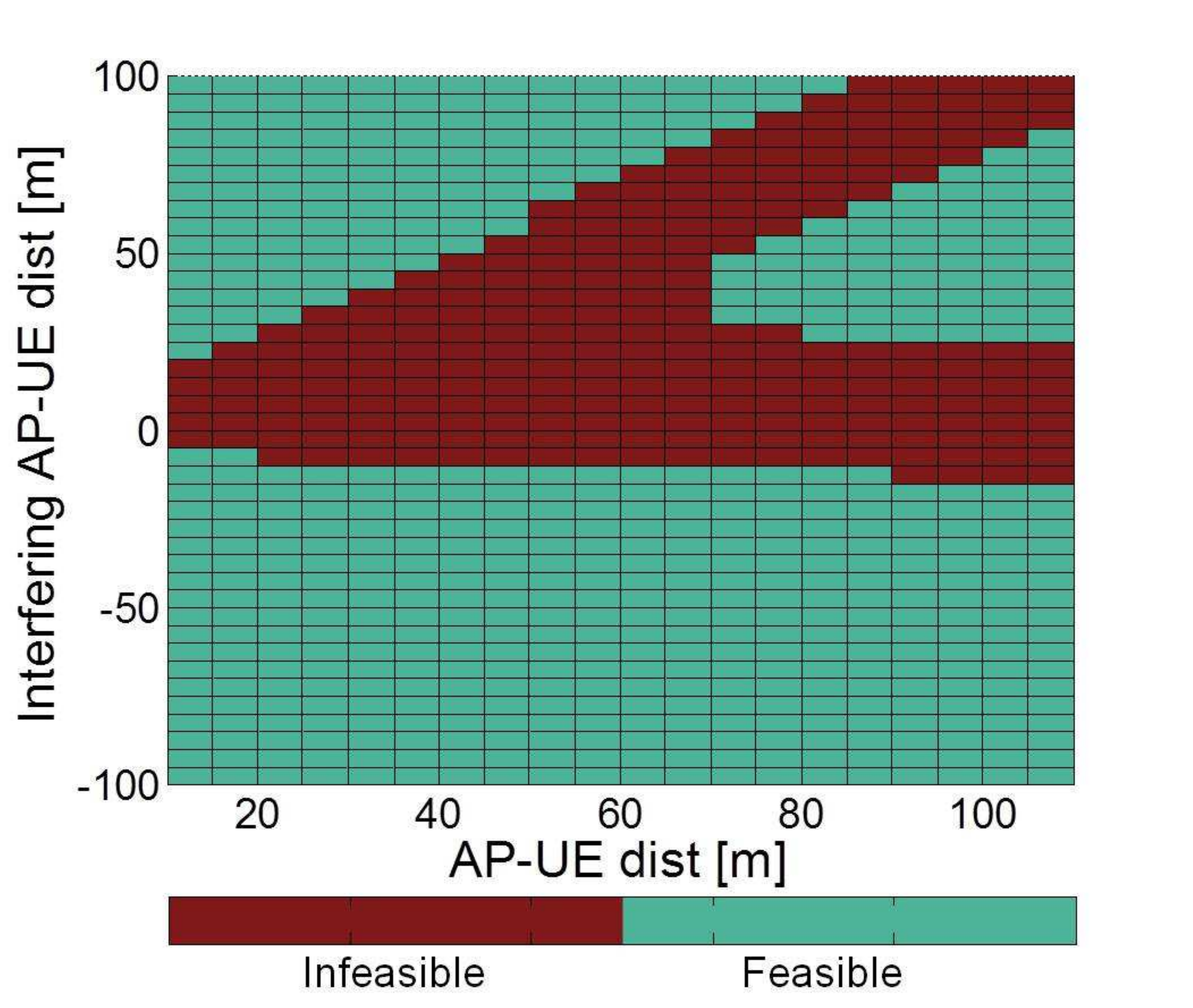}
    \label{fig:w_opt_feasible}
}
\subfigure[A heat map of WiFi throughput when time division channel access coordination (Mbps)]{
	\includegraphics[height=1.75in,width=2.3in]{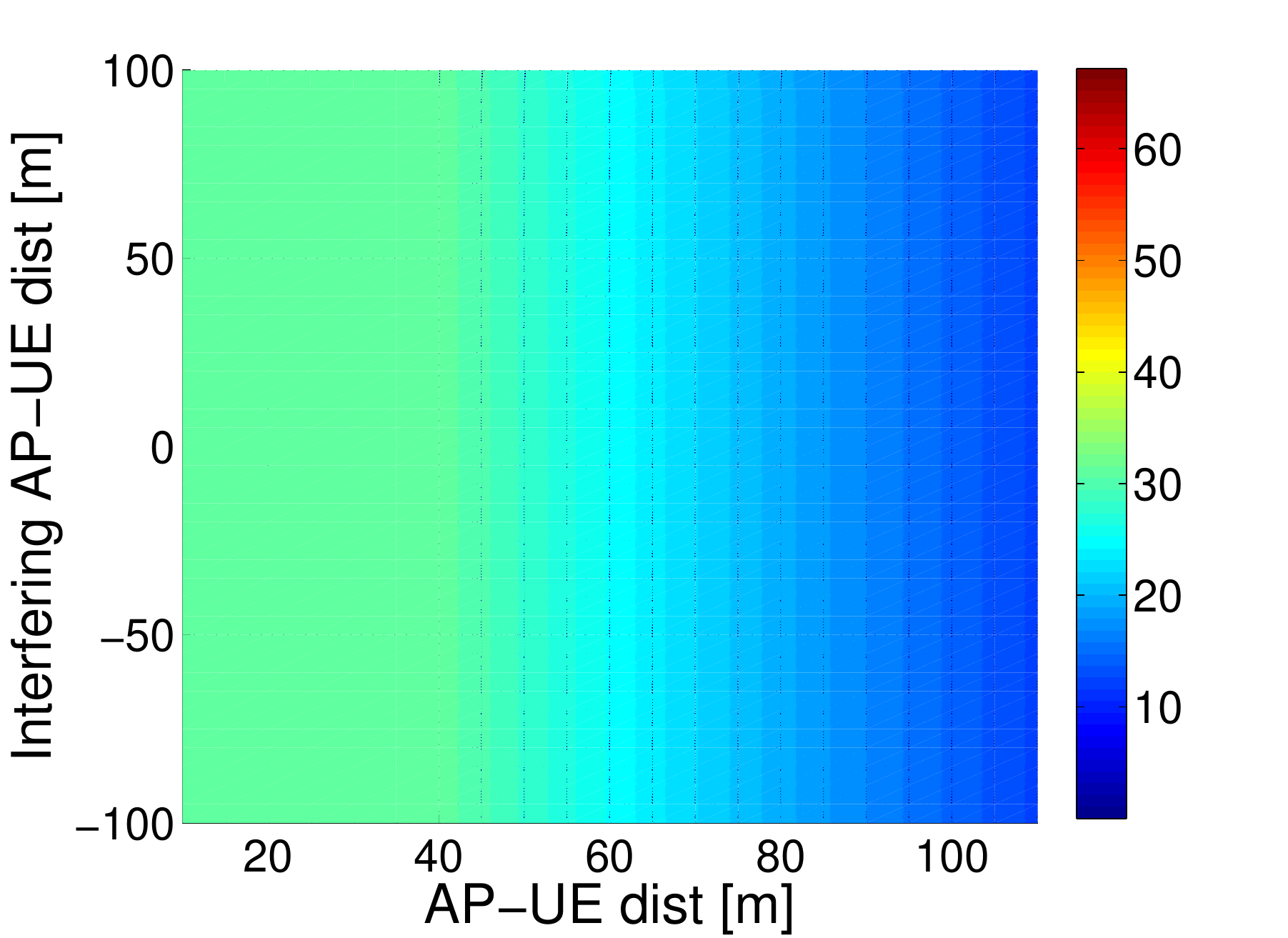}
    \label{fig:W_wiL_pwrTDOpt_3d}
}
\caption[Optional caption for list of figures]{WiFi performance under joint WiFi and LTE power control optimization  
}
\label{fig:wifi_Opt}
\end{center}
\end{figure*}

\begin{figure*}[t]
\begin{center}
\subfigure[A heat map of LTE throughput when joint power Coordination (Mbps)]{
	\includegraphics[height=1.75in,width=2.3in]{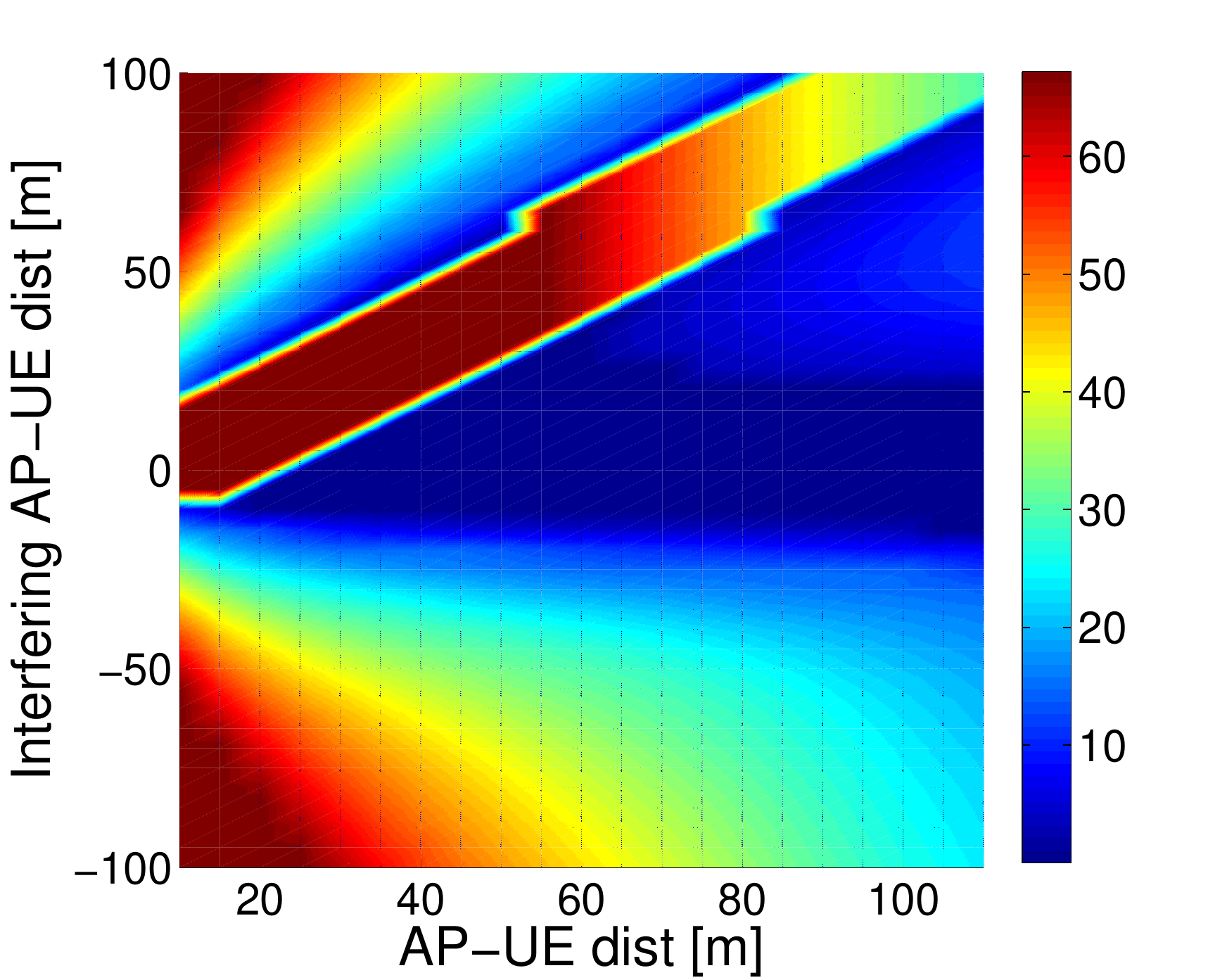}
    \label{fig:L_wiL_pwrOpt_3d}
}
\subfigure[Feasibility region of joint power Coordination]{
	\includegraphics[height=1.75in,width=2.1in]{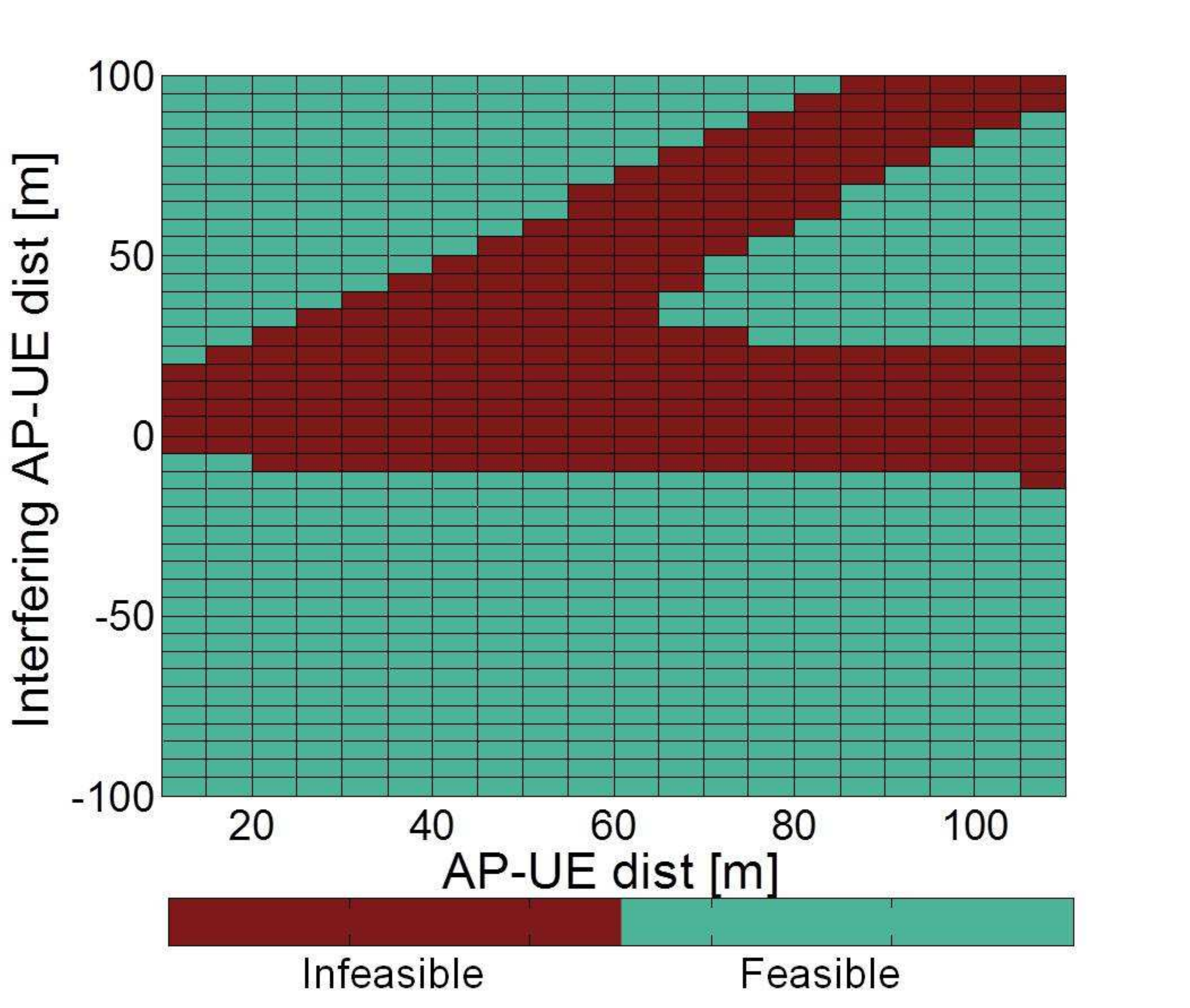}
    \label{fig:L_opt_feasible}
}
\subfigure[A heat map of LTE throughput when time division channel access coordination (Mbps)]{
	\includegraphics[height=1.75in,width=2.3in]{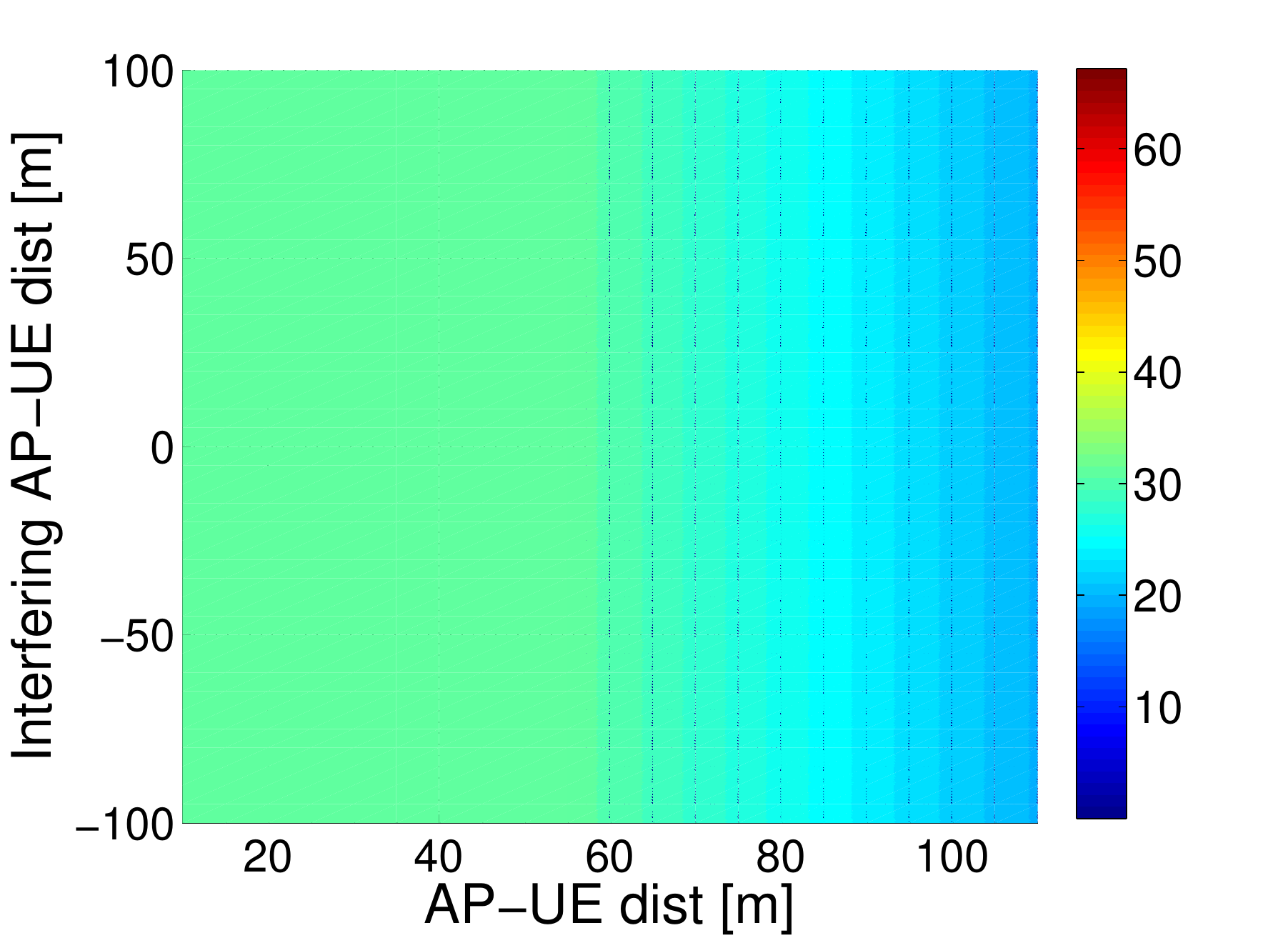}
    \label{fig:L_wiL_pwrTDOpt_3d}
}
\caption[Optional caption for list of figures]{WiFi performance under joint WiFi and LTE power control optimization  
}
\label{fig:lte_Opt}
\end{center}
\end{figure*}

\subsubsection{Power control optimization across network of same RAT}
Based on the GP-formulation, the transmission power of the APs across the same network $i, i\in\{w,l\},$ are optimized such that
\begin{equation}
\begin{aligned}
& {\text{maximize}}
& & \sum_{i\in\mathcal{W}} R_i\\
& \text{subject to}
& & R_i \geq R_{i,\min}, \;\; i\in\mathcal{W}\\
&&& 0 \leq P_i \leq P_{\max}, \;\; i\in\mathcal{W},\\
&&& {\sum_{k\in M^b_i} P_kG_{ik} + N_0} < \lambda_c, \; i\in\mathcal{W}.\\
\end{aligned}
\end{equation}
and
\begin{equation}
\begin{aligned}
& {\text{maximize}}
& & \sum_{i\in\mathcal{L}} R_i\\
& \text{subject to}
& & R_i \geq R_{i,\min}, \;\; i\in\mathcal{L}\\
&&& 0 \leq P_i \leq P_{\max}, \;\; i\in\mathcal{L}.\\
\end{aligned}
\end{equation}
Here, the objective function is equivalent to maximizing the product of SINRs at the networks $i, i\in \{w,l\}$. The first and second constraints ensure that we meet the minimum SINR and transmission power limits requirements at all links of $i$. In this formulation, SINR at WiFi and LTE respectively given as
\[
\begin{aligned}
S_i =& \frac{P_iG_{ii}}{N_0},\;\; i\in\mathcal{W},\\
S_i =& \frac{P_iG_{ii}}{\sum_{j\in\mathcal{L}, j \neq i}P_jG_{ij} + N_0},\;\; i\in\mathcal{L}.
\end{aligned}
\]
which are first cases in equations (\ref{eq:w_sinr}) and (\ref{eq:l_sinr}), respectively.

\subsubsection{Joint time division channel access optimization}
This is the joint optimization across both WiFi and LTE networks which is formulated as given below
\begin{equation}
\begin{aligned}
& {\text{maximize}}
& & \min\left(\eta R_{i\in\mathcal{W}},(1-\eta)R_{j\in\mathcal{L}}\right)\\
& \text{subject to}
& & 0 \leq \eta \leq 1.\\
\end{aligned}
\end{equation}

\eat{\begin{equation}
\begin{aligned}
& {\text{maximize}}
& & \min\left(\eta R^w_{i\in W},(1-\eta)R^l_{j\in L}\right)\\
& \text{subject to}
& & \delta_1 \leq \frac{\sum_{i\in W}R^w_i}{\sum_{j\in L}R^l_j} \leq \delta_2,\\
&&& 0 \leq \eta \leq 1.\\
\end{aligned}
\end{equation}}
Here, throughput values at all WiFi and LTE nodes are considered as a constant, which is the output of the previous step. Time division channel access parameter $\eta$ is optimized so that it maximizes the minimum throughput across all UEs.

\eat{
\begin{itemize}
\item{$N \rightarrow$}
number of each WiFi and LTE links,
\item{$w, l \rightarrow$}
indices for the WiFi and LTE network, respectively;
\item{$W \rightarrow$}
the set of WiFi links
\item{$L \rightarrow$}
the set of LTE links
\item{$C^w \rightarrow$}
boolean matrix $C^w \in \{0,1\}^{N\times K}$ is the channel assignment matrix of WiFi where $C^w_{ik} = 1$ if WiFi $i$ operates on channel $k$, $C^w_{ik} = 0$ otherwise,
\item{$C^l \rightarrow$}
boolean matrix $C^l \in \{0,1\}^{N\times K}$ is the channel assignment matrix of LTE where $C^l_{ik} = 1$ if LTE $i$ operates on channel $k$, $C^l_{ik} = 0$ otherwise,
\item{$P^l_i \rightarrow$}
transmission power of LTE $i$, where $0 \leq P^l_i \leq P^l_{\max}$,
\item{$G_{ij} \rightarrow$}
channel gain between nodes $i$ and $j$,
\item{$N_0 \rightarrow$}
noise density of the channel,
\item{$\gamma^j_i \rightarrow$}
Signal-to-Interference-plus-Noise-Ratio (SINR) at $i$ of network $j \in {w,l}$,
\item{$M^a \rightarrow$}
boolean matrix $M^a \in \{0,1\}^{N\times N}$ represents WiFi CSMA matrix where $M^a_{ij} = 1$ if WiFi APs $i$ and $j$ are in CSMA range of each other, $M^a_{ij} = 0$ otherwise, (convention: $M^a_{ij} = 1, i = j$),
\item{$a_{i} \rightarrow$}
approximate reduction in channel access time fraction at WiFi $i$ due to sharing of channel with other WiFis in the CSMA range
which are on the same channel as $i$
\item{$M^b \rightarrow$}
boolean matrix $M^b \in \{0,1\}^{N\times N}$ represents WiFi interference matrix where $M^b_{ij} = 1$ if WiFi APs $i$ and $j$ are in interference range of each other, $M_{ij} = 0$ otherwise,
\item{$b_{i} \rightarrow$}
reduction in channel access time fraction due to hidden node interference (interference from APs outside the carrier sense but strong enough signal to affect ongoing transmission)
\item{$Z^j_i \rightarrow$}
throughput at $i$ of network $j \in {w,l}$,
\end{itemize}

Parameters are:
\begin{itemize}
\item{$\lambda_c \rightarrow$}
the clear channel assessment (CCA) threshold for WiFi,
\item{$P^w \rightarrow$}
transmission power of WiFi (fixed value),
\item{$\beta \rightarrow$}
$\beta \in [0,1]$ captures  the average effect of hidden node interference per interferer,
\item{$\eta_B \rightarrow$}
LTE bandwidth efficiency due to factors adjacent channel leakage ratio and practical filter, cyclic prefix, pilot assisted channel estimation, signaling overhead
\item{$\gamma^j_{\min} \rightarrow$}
Minimum SINR required for network $j \in {w,l}$,
\item{$\gamma^j_{\max} \rightarrow$}
Maximum SINR required for network $j \in {w,l}$,
\item{$P^l_{\max} \rightarrow$}
Maximum transmission power allowed for LTE,
\item{$\zeta_{\min},\zeta_{\max} \rightarrow$}
proportional fairness constants to balance WiFi and LTE throughput, ideally $\zeta_{\min} = \zeta_{\max} = 1$.
\end{itemize}

\textbf{Objective}: to maximize sum-throughput across WiFi and LTE network\\
\textbf{Controlling variables}: $C^w, C^l$ and $P^l_i, \forall i\in L$

\begin{equation}
\begin{aligned}
& {\text{maximize}}
& & \sum_{i\in W} Z^w_i + \sum_{j\in L} Z^l_j\\
& \text{subject to}
& & C^w_{ik} \in \{0,1\}, \;\; \forall i\in W,\\
&&& C^l_{ik} \in \{0,1\}, \;\; \forall i\in L,\\
&&& \sum_{k} C^w_{ik} = 1, \; \forall i\in W,\\
&&& \sum_{k} C^l_{ik} = 1, \; \forall i\in L,\\
&&& {\sum_{k}\sum_{j\in L} C^w_{ik} C^l_{jk} P^l_{j}G_{ij}} + N_0 < \lambda_c, \; \forall i\in W,\\
&&& \gamma^l_i = \frac{P^l_iG_{ii}}{\sum_{k}C^l_{ik}\left( \sum_{m\in L, i\neq m} C^l_{mk}P^l_m G_{im} + \sum_{n\in W} C^w_{nk}a_nP^w G_{in} \right) + N_0}, \;\; \forall i\in L,\\
&&& \gamma^w_i = \frac{G_{ii}P^w}{\sum_{k}\sum_{j\in L} C^w_{ik} C^l_{jk} P^l_j G_{ij} + N_0}, \;\; \forall i\in W,\\
&&& a_{i} = \frac{1}{\sum_{k}\sum_{j\in W} C^w_{ik} C^w_{jk} M^a_{ij}}, \;\;  \forall i\in W,\\
&&& b_i = \frac{1}{1+\beta\sum_{k}\sum_{j\in W} C^w_{ik} C^W_{jk} M^b_{ij}}, \;\;  \forall i\in W,\\
&&& Z^w_i = a_ib_if(\gamma^w_i), \; \forall i\in W;\\
&&& Z^l_i = \eta_B\log(1 + \gamma^l_i), \; \forall i\in L,\\
&&& \zeta_{\min} \leq \frac{\sum_{i\in W} Z^w_i}{\sum_{j\in L} Z^l_j} \leq \zeta_{\max}, \\
&&& \gamma^w_{\min} \leq \gamma^w_i \leq \gamma^w_{\max}, \;\; \forall i\in W,\\
&&& \gamma^l_{\min} \leq \gamma^l_i \leq \gamma^l_{\max}, \;\; \forall i\in L,\\
&&& 0 \leq P^l_i \leq P^l_{\max}, \;\; \forall  i\in L.\\
\end{aligned}
\end{equation}
}

\section{Evaluation of Joint Coordination}
\label{sec:results}

\subsection{Single Link Co-channel Deployment}
We begin with the motivational example of co-channel deployment of one Wi-Fi and one LTE links, as described in \S~\ref{subsec:motiEx}. Figure~\ref{fig:wifi_Opt} shows the heatmap of improved throughput of Wi-Fi link, when joint Wi-Fi and LTE coordination is provided in comparison with the throughput with no coordination as shown in figure~\ref{fig:wifi_noOpt} . Similarly, figure~\ref{fig:lte_Opt} shows the heatmap of improved throughput of LTE link, when joint coordination is provided in comparison with the throughput with no coordination, as shown in figure~\ref{fig:lte_noOpt}.

For both the figures~\ref{fig:wifi_Opt} and \ref{fig:lte_Opt}, in their respective scenarios, though joint power control improves the overall throughput for most of topological scenarios (see Figure (a) of \ref{fig:wifi_Opt} and \ref{fig:lte_Opt}), it is not an adequate solution for topological combination marked by \textit{infeasible} region as given in figure (b) of \ref{fig:wifi_Opt} and \ref{fig:lte_Opt}. The infeasible region signifies the failure to attain the CCA threshold at Wi-Fi AP and link SINR requirement when the UE and interfering AP are very close to each other. When we apply time division channel access optimization for a given scenario, we do not observe any infeasible region, in fact optimization achieves almost equal and fair throughput at both Wi-Fi and LTE link, as shown in figure~(c) of \ref{fig:wifi_Opt} and \ref{fig:lte_Opt}. On the downside, this optimization does not consider cases when Wi-Fi and LTE links can operate simultaneously without causing severe interference to each other. In such cases, throughput at both Wi-Fi and LTE get degraded.
\begin{figure}[t]
\begin{center}
\includegraphics[height=2.15in,width=3.5in]{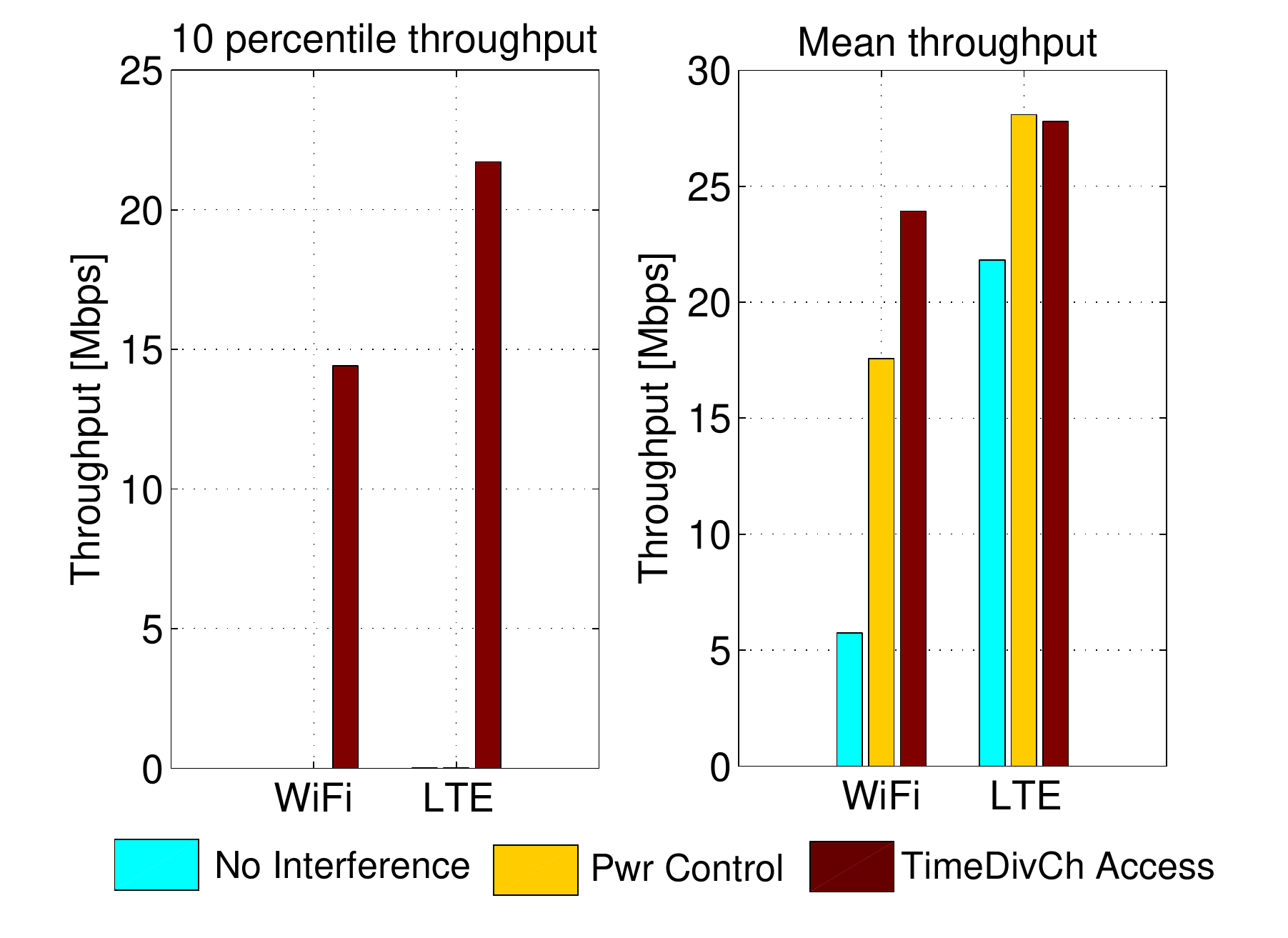}
\caption{10 percentile and mean LTE throughput for a single link WiFi and LTE co-channel deployment}
\label{fig:oneLW_bar10Mean}
\end{center}
\end{figure}

Figure~\ref{fig:oneLW_bar10Mean} summarizes the performance of Wi-Fi and LTE links in terms of 10 percentile and mean throughput. We note that the 10 percentile throughput of both Wi-Fi and LTE is increased to ~ $15-20$ Mbps for time division coordination compared to $\sim$ zero throughput for no and power coordination. We observe $~200\%$ and $~350\%$ Wi-Fi mean throughput gains due to power and time division channel access, respectively, compared to no coordination.  For LTE, throughput gains for both of these coordination is $\sim 25-30\%$. It appears that time division channel access coordination does not offer any additional advantage to LTE in comparison with power coordination. But it brings the throughput fairness between Wi-Fi and LTE which is required for the co-existence in the shared band.

\subsection{Multiple Links Co-channel Deployment}
Multiple overlapping Wi-Fi and LTE links are randomly deployed in 200-by-200 sq. meter area which depicts the typical deployment in residential or urban hotspot. The number of APs of each Wi-Fi and LTE networks are varied between 2 to 10 where number of Wi-Fi and LTE links are assumed to be equal. For the simplicity purpose, we assume that only single client is connected at each AP and their association is predefined. The given formulation can be extended for multiple client scenarios. In the simulations, the carrier sense and interference range for Wi-Fi devices are set to 150 meters and 210 meters, respectively. The hidden node interference parameter is set to 0.25.

\begin{figure}[t]
\begin{center}
\subfigure[10 percentile and mean Wi-Fi throughput for $N =\{2,5,10\}$]{
	\includegraphics[height=2.15in,width=3.5in]{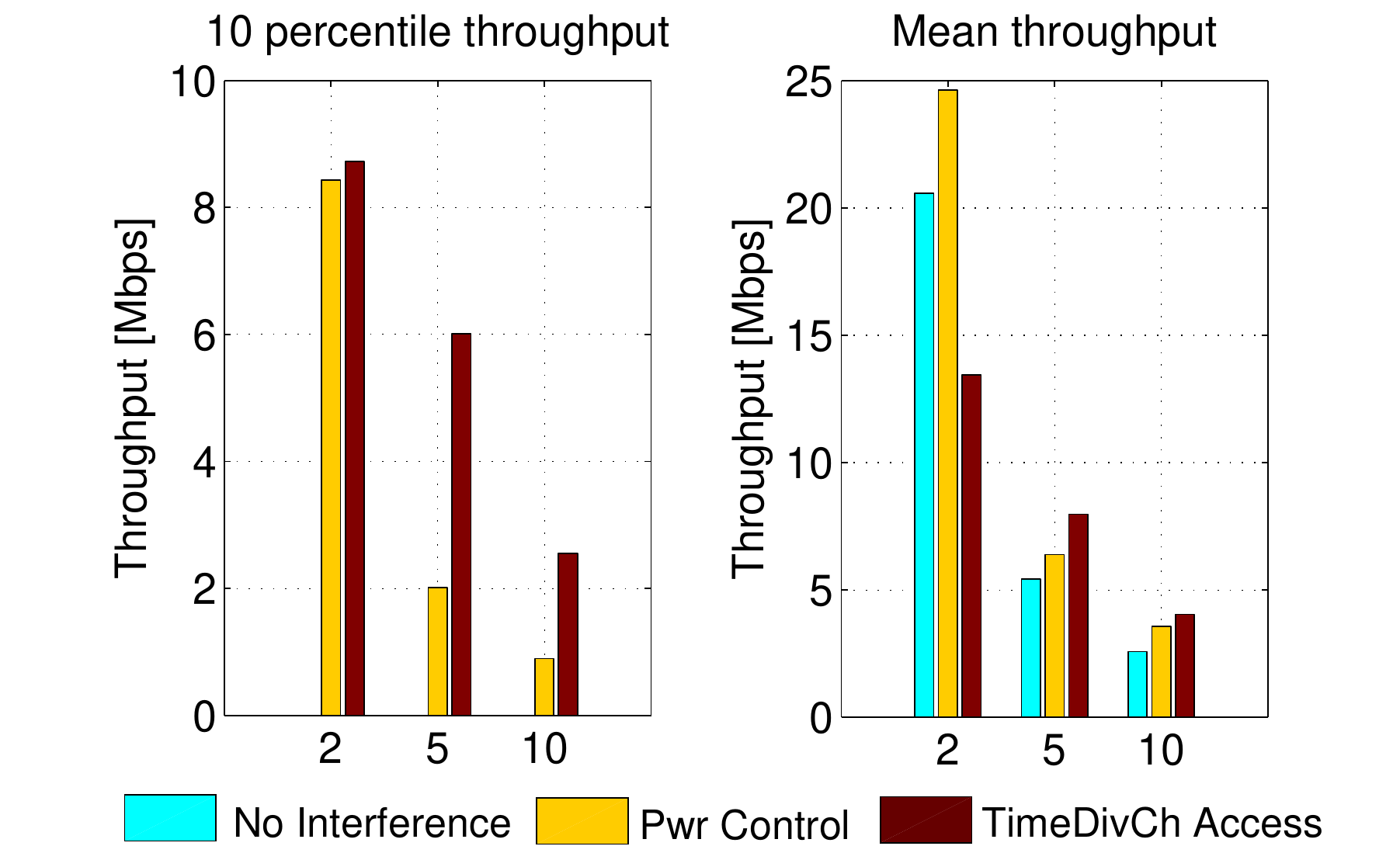}
    \label{fig:w_multiLink}
}
\subfigure[10 percentile and mean LTE throughput for $N =\{2,5,10\}$]{
	\includegraphics[height=2.35in,width=3.5in]{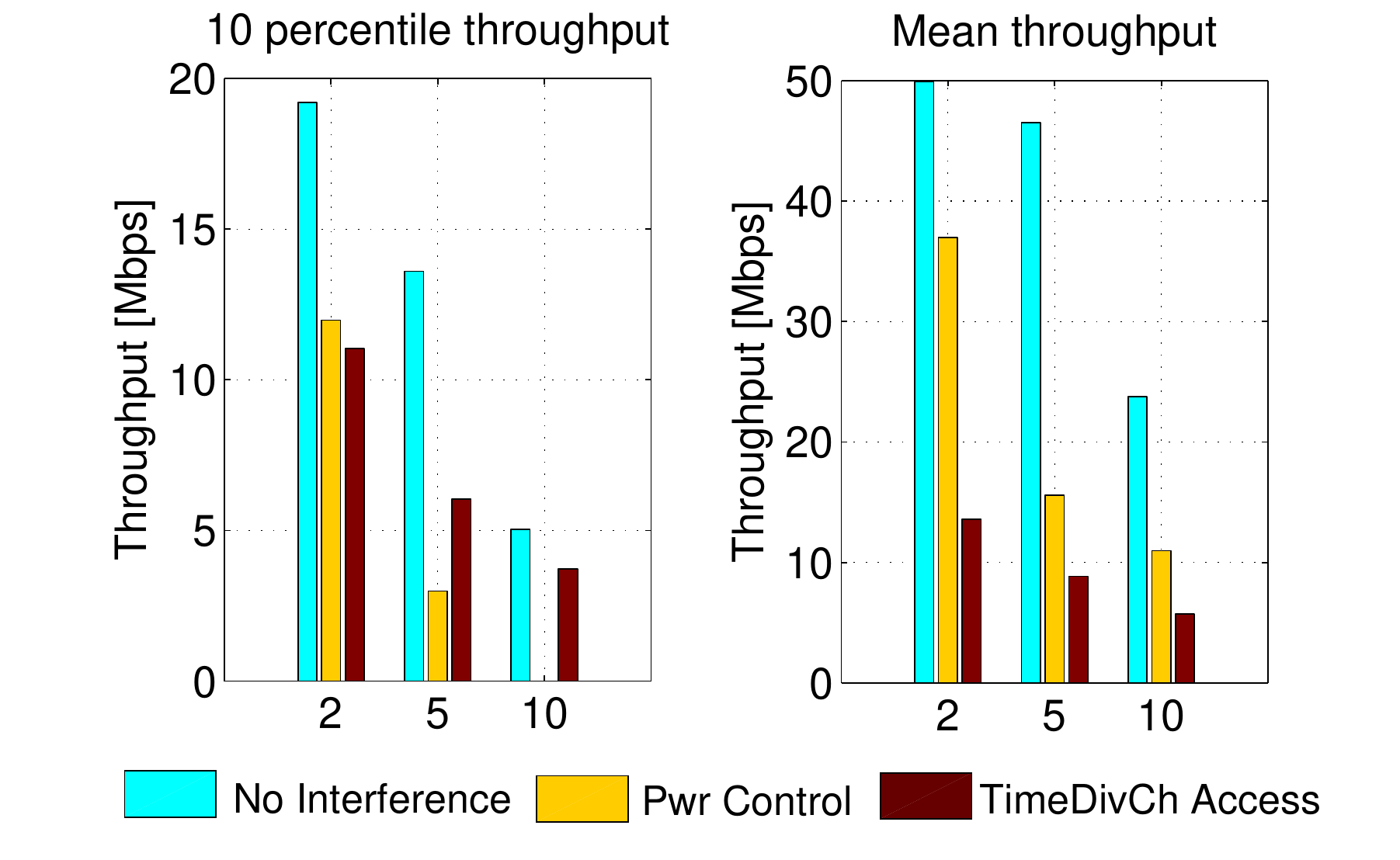}
    \label{fig:L_multiLink}
}
\caption[Optional caption for list of figures]{Multi-link throughput performance under power control and time devision channel access optimization. $N = $ no. of LTE links = no. of Wi-Fi links. 
}
\label{fig:wifi_Opt_1}
\end{center}
\end{figure}

Figure~\ref{fig:w_multiLink} and \ref{fig:w_multiLink} show the percentile and mean throughput values of Wi-Fi and LTE links, respectively, for when number of links for each Wi-Fi and LTE networks is set at $N = \{2,5,10\}$. The throughput performance is averaged over 10 different deployment topologies of Wi-Fi and LTE links. From figure~\ref{fig:w_multiLink}, it is clear that 10 percentile Wi-Fi UEs get throughput starved due to LTE interference with no coordination. This is consistent with results from single link simulations. With coordination, both joint power control and time division channel access, we achieve a large improvement in the 10 percentile throughput. Joint power control improves mean Wi-Fi throughput by 15-20$\%$ for all $N$. On the  other hand, time division channel access achieves throughput gain (40-60$\%$) only at higher values of $N = \{5,10\}$.

Throughput performance of LTE, on the other hand, get deteriorates in the presence of coordination compared to when no coordination is provided. This comes from the fact that, in case of no coordination, LTE causes undue impact at Wi-Fi which makes them to hold off data transmission and LTE experiences no Wi-Fi interference. The joint coordination between Wi-Fi and LTE networks brings the notion of fairness in the system and allocates spectrum resources to suffered Wi-Fi links. In the joint power control optimization, though certain LTE links (maximum 1 link for $N = 10$) have to be dropped from network with zero throughput, the overall mean throughput is greater than $150$ to $400\%$ than Wi-Fi throughput.

We observe that for small numbers of Wi-Fi links, joint time division channel access degrades the performance of both Wi-Fi and LTE. But as number of links grows, coordinated optimization results in allocation of orthogonal resources (e.g. separate channels) gives greater benefit than full sharing of the same spectrum space, as is the case for power control optimization.

%

\section{Conclusion}
\label{sec:conclude}
This paper investigates inter-system interference in shared spectrum scenarios with both Wi-Fi and LTE in the same band.  An analytical model has been developed for evaluation of the performance and the model has been partially verified with experimental data. The results show that significant performance degradation results from uncoordinated operation of Wi-Fi and LTE in the same band. To address this problem, we further presented an architecture for coordination between heterogeneous networks, with a specific focus on  LTE-U and Wi-Fi, to cooperate and coexist in the same area. This framework is used to exchange information between the two networks for a logically centralized optimization approach that improves the aggregate throughput of the network. Our results show that, with joint power control and time division multiplexing, the aggregate throughput of each of the networks becomes comparable, thus realizing fair access to the spectrum. 

\noindent\textbf{Acknowledgment}: Research is supported by NSF EARS program- grant CNS-1247764.

\bibliographystyle{IEEEbib}
\small
\bibliography{ref,ref_5G}

\end{document}